\documentclass[12pt]{article}
\usepackage{amssymb,amsmath,amscd}
\usepackage{epsfig}
\usepackage{epstopdf}
\usepackage{latexsym}
\usepackage{graphicx}
\usepackage{subfigure}
\usepackage{booktabs}
\usepackage{bbm}
\usepackage[margin=15pt,small]{caption}
\usepackage{enumitem}
\usepackage[toc]{appendix}
\usepackage[numbers,compress]{natbib}

\usepackage{color}
\usepackage{datetime}
\usepackage[
      colorlinks=false,
      linkcolor=darkblue,  
      urlcolor=blue,    
      filecolor=blue,     
      citecolor=red,
linktocpage=true,
      pdfstartview=FitV,
      bookmarksopen=true    
      ]{hyperref}

\setlist{nolistsep}

\ifpdf
\fi

\let\oldbibliography\thebibliography
\renewcommand{\thebibliography}[1]{\oldbibliography{#1}
\setlength{\itemsep}{0pt}} 

\newcommand*{\boxedcolor}{red}
\makeatletter
\renewcommand{\boxed}[1]{\textcolor{\boxedcolor}{%
  \fbox{\normalcolor\m@th$\displaystyle#1$}}}
\makeatother

\definecolor{cardinal}{rgb}{0.6,0,0}
\definecolor{darkgreen}{rgb}{0,0.5,0}
\definecolor{golden}{rgb}{0.92, 0.7, 0}
\definecolor{midnight}{rgb}{0, 0, 0.5}
\definecolor{darkblue}{rgb}{0.2, 0, 0.8}

\def\coeff#1#2{\relax{\textstyle {#1 \over #2}}\displaystyle}
\def\ds{\displaystyle}

\def\cA{{\cal A}}

\def\cK{{\cal K}}

\def\cN{{\cal N}}

\def\cP{{\cal P}}

\def\cT{{\cal T}}

\def\eql{=}

\def\eql{=}

\def\Tr{{\rm Tr}\,}

\def\RR{\mathbb{R}}

\def\cL{{\cal L}}

\def\fg{\frak{g}}

\def\fa{\frak{a}}

\def\bfs#1{{\boldsymbol{#1}}}

\def\cals#1{\mathcal{#1}}

\def\cA{{\cals A}}

\def\xxi{z}
\def\R{{$R$}}

\def\varphial{\varphi_\alpha }
\def\varphibe{\varphi_\beta }
\def\varphichi{\varphi_\chi }
\def\varphih{\varphi_h }

\newcommand{\sect}[1]{\setcounter{equation}{0}\section{#1}}
\renewcommand{\theequation}{\arabic{section}.\arabic{equation}}

\begin{document}  

\begin{titlepage}
 
\medskip
\begin{center} 
{\Large \bf  $\bfs{(0,2)}$ SCFTs from the Leigh-Strassler Fixed Point}

\bigskip
\bigskip
\bigskip
\bigskip

{\bf Nikolay Bobev,${}^{(1)}$  Krzysztof Pilch,${}^{(2)}$  and Orestis Vasilakis${}^{(2)}$ \\ }
\bigskip
${}^{(1)}$
Perimeter Institute for Theoretical Physics \\
31 Caroline Street North, Waterloo, ON N2L 2Y5, Canada
\vskip 5mm
${}^{(2)}$ Department of Physics and Astronomy \\
University of Southern California \\
Los Angeles, CA 90089, USA  \\
\bigskip
nbobev@perimeterinstitute.ca,~pilch@usc.edu,~vasilaki@usc.edu  \\
\end{center}

\bigskip
\bigskip
\bigskip
\bigskip

\begin{abstract}

\noindent  
\end{abstract}

\noindent We show that there is a family of two-dimensional $(0,2)$ SCFTs associated with twisted compactifications of the four-dimensional $\mathcal{N}=1$ Leigh-Strassler fixed point on a closed hyperbolic Riemann surface. We calculate the central charges for this class of theories using anomalies and $c$-extremization. In  a suitable truncation of the five-dimensional maximal supergravity, we construct supersymmetric $AdS_3$ solutions that are holographic duals of those two-dimensional $(0,2)$ SCFTs.  We also exhibit supersymmetric  domain wall solutions that are holographically dual  to the RG flows between the four-dimensional and two-dimensional theories.

\vfill

\end{titlepage}


\newpage

\setcounter{tocdepth}{2}
\tableofcontents

\section{Introduction}

It has become clear in recent years that one can engineer large classes of three- and four-dimensional superconformal field theories (SCFTs) by studying twisted compactifications of the six-dimensional $(2,0)$ theory on three-manifolds and Riemann surfaces \cite{Gaiotto:2009we,Dimofte:2011ju}. The higher-dimensional theory on the one side, and the geometry of the internal manifold on the other, allow one to uncover dualities in the lower-dimensional SCFTs. In addition, one is led to various relations between correlations functions in the SCFTs and a topological field theory on the compactification manifold. 

In view of this rich structure, it is natural to extend these ideas to two-dimensional SCFTs. One possible approach is to study twisted compactifications of the six-dimensional $(2,0)$ theory on four-manifolds preserving at least $(0,2)$ supersymmetry in the two remaining directions. This has been attempted in \cite{Benini:2013cda,Gadde:2013sca,Gadde:2013lxa} but, due to the small amount of supersymmetry, the complicated geometry of four-manifolds, and the limited understanding of the $(2,0)$ theory, there have been only a few quantitative results. A simpler and potentially more fruitful strategy to get a handle on $(0,2)$ SCFTs is to study twisted compactifications of four-dimensional supersymmetric theories on a Riemann surface. This idea has been explored to some extent for four-dimensional $\mathcal{N}=4$ SYM in \cite{Bershadsky:1995vm,Maldacena:2000mw}, where some particular SCFTs with $(4,4)$ and $(2,2)$ supersymmetry were studied. Recently, there has been an extension of these constructions to theories with $(0,2)$ supersymmetry obtained either from $\mathcal{N}=4$ SYM \cite{Benini:2013cda,Benini:2012cz,Almuhairi:2011ws,Donos:2011pn}, or various $\mathcal{N}=1$ theories in four dimensions \cite{BBC,Kutasov:2013ffl,Kutasov:2014hha}. The goal of this paper is to enlarge further the class of such two-dimensional $(0,2)$ SCFTs by studying a twisted compactification of the four-dimensional $\mathcal{N}=1$ Leigh-Strassler (LS) SCFT, in the following referred to as the LS fixed point, \cite{Leigh:1995ep}. 

For our purposes it will be most useful to think of the LS fixed point as a strongly coupled $\mathcal{N}=1$ SCFT obtained from $\mathcal{N}=4$ SYM by an RG flow induced by turning on a mass for one of the three adjoint chiral superfields \cite{Leigh:1995ep}. The theory has an $U(1)$ $R$-symmetry and a $SU(2)$ flavor symmetry inherited from the $SO(6)$ $R$-symmetry of $\mathcal{N}=4$ SYM. To obtain a supersymmetric two-dimensional theory we place the four-dimensional SCFT on $\mathbb{R}^{1,1}\times \Sigma_{\mathfrak{g}}$, where $\Sigma_{\mathfrak{g}}$ is a closed Riemann surface of genus $\mathfrak{g}$, and turn on a background gauge field for an Abelian subgroup of this $U(1)_R \times SU(2)_F$ global symmetry. Guided by the previous work in \cite{Benini:2013cda,Bershadsky:1995vm,Maldacena:2000mw,Benini:2012cz,Almuhairi:2011ws,Donos:2011pn}, we  assume that the low-energy two-dimensional theory is conformal and then use the knowledge of the 't Hooft anomalies of the LS fixed point as well as two-dimensional  $c$-extremization \cite{Benini:2013cda,Benini:2012cz} to calculate the left and right Virasoro central charge of the two-dimensional $(0,2)$ IR fixed point. We find that for all possible topological twists on a hyperbolic Riemann surface, the central charges are positive, which is compatible with unitary and suggests that the fixed points indeed exist. However, with only $(0,2)$ supersymmetry in two-dimensions, it is typically difficult to establish rigorously the existence of a fixed point in the IR.\footnote{This is reminiscent of the situation in four-dimensional $\mathcal{N}=1$ theories \cite{Seiberg:1994pq}.} One way to obtain more evidence for the existence of the IR fixed points is to construct string theory or supergravity backgrounds which are holographic duals to the field theories of interest. This is one of the tools we will utilize in our work. To this end we find new supersymmetric $AdS_3 \times \Sigma_{\mathfrak{g}}$ supergravity solutions in the spirit of \cite{Maldacena:2000mw,Benini:2012cz,Benini:2013cda,Bah:2011vv,Bah:2012dg}. In addition, we construct numerical solutions which we interpret as holographic RG flows from an asymptotically locally $AdS_5$ solution to the $AdS_3 \times \Sigma_{\mathfrak{g}}$ vacua of interest. This shows that, at least in the regime of validity of holography, the RG flows are realized dynamically.

The paper is organized as follows. In Section~\ref{Sec:FieldTheory} we  summarize the salient features of the $\mathcal{N}=1$ LS SCFT. We then put the theory on $\mathbb{R}^{1,1}\times \Sigma_{\mathfrak{g}}$, perform a partial topological twist and  calculate the conformal anomaly of the resulting family of two-dimensional $(0,2)$ SCFTs. In Section \ref{Sec:SUGRAtrunc} we   present the truncation of five-dimensional supergravity, which we use in Section \ref{Sec:AdS3sol} to construct the family of supersymmetric $AdS_3\times \Sigma_{\frak{g}}$ vacua dual to the $(0,2)$ SCFTs of interest. In Section \ref{Sec:HoloRG} we   construct holographic RG flows which interpolate between the gravity dual of the $\mathcal{N}=1$ LS fixed point and the supersymmetric $AdS_3\times \Sigma_{\frak{g}}$ solutions found in Section~\ref{Sec:AdS3sol}. In addition, we construct holographic RG flows connecting the same $AdS_3\times \Sigma_{\frak{g}}$ vacua and the maximally supersymmetric $AdS_5$ solution. We conclude in Section \ref{Sec:Conclusions} with some comments and open questions. In Appendix~\ref{appendixA} we  provide some details on the supergravity truncation used in the paper and in Appendix \ref{appendixB} we discuss the correspondence between the Chern-Simons couplings in five- and three-dimensional supergravity and the anomalies for global symmetries in the dual field theory.

\sect{Field theory}
\label{Sec:FieldTheory}
 
It was shown in \cite{Benini:2012cz,Benini:2013cda}, following the earlier work  \cite{Bershadsky:1995vm,Maldacena:2000mw}, that there is a rich family of $(0,2)$ SCFTs in two dimensions obtained by compactifying $\mathcal{N}=4$ SYM on a Riemann surface and flowing to the IR. It is also well known that there is an interacting four-dimensional $\mathcal{N}=1$ SCFT, known as the LS fixed point \cite{Leigh:1995ep}, which can be obtained by integrating out one of the three adjoint chiral superfields in $\mathcal{N}=4$ SYM. The goal of this section is to provide some evidence that, when the LS fixed point is put on $\mathbb{R}^{1,1}\times \Sigma_{\mathfrak{g}}$ with a partial topological twist, the effective two-dimensional theories in the IR are a family of $(0,2)$ SCFTs similar to the ones studied in \cite{Benini:2012cz,Benini:2013cda}. We will argue in favor of the existence of these fixed points using 't Hooft anomaly matching together with $c$-extremization \cite{Benini:2012cz,Benini:2013cda}.

\subsection{Anomalies}
\label{subsec:ANSYM}

Recall that $\cN=4$ SYM  can be viewed as an $\mathcal{N}=1$ theory of a vector multiplet with gauge field $A_{\mu}$ and gaugino  $\lambda$, coupled to  three adjoint chiral multiplets, $\Phi_i$, each containing a complex scalar, $\phi_i$, as its lowest component and a complex fermion, $\chi_i$. The $\cN=1$ superpotential reads
\begin{equation}
W = \Tr \Phi_1\Phi_2\Phi_3\;.
\end{equation}
In this formulation only an  $SU(3)\times U(1)_R^{\mathcal{N}=4}$  subgroup of the  $SO(6)\simeq SU(4)$ \R-symmetry of  the $\mathcal{N}=4$ theory is manifest. The $U(1)_R^{\mathcal{N}=4}$ is the superconformal \R-symmetry  generated by
\begin{equation}\label{TRN=4}
T_R^{\mathcal{N}=4} = \frac{2}{3}\left(T_1+T_2+T_3\right)\;,
\end{equation}
where the  three $T_i$'s  are the generators  of the Cartan subalgebra of  $SO(6)$ given by the three diagonal $SO(2)_i$'s.  The charges of the four adjoint fermions with respect to those $SO(2)_i$'s and  the resulting  \R-charges are given in Table~\ref{table:one}. The \R-charges of the complex scalars, $\phi_i$, in the three chiral multiplets are all the same and 
equal to $2/3$.
By evaluating the usual triangle diagrams with the chiral fermions, one finds that the cubic and linear 't Hooft anomalies for   $U(1)_R^{\mathcal{N}=4}$  are given by:
\begin{equation}\label{thooftn4}
\begin{split}
\mathop{\rm tr} R^3_{\mathcal{N}=4} &\eql  d_{G}\,\Big [1^3-3\Big({1\over 3}\Big)^3\,\Big] = \frac{8}{9}\,d_G\;, \qquad 
\mathop{\rm tr} R_{\mathcal{N}=4}  \eql  d_{G}\,\Big [1-3\,\Big({1\over 3}\Big)\,\Big] = 0\;,
\end{split}
\end{equation}
where $d_G$ is the dimension of the gauge group ($d_G=N^2-1$ for $SU(N)$). 

The superconformal  \R-symmetry current in any $\mathcal{N}=1$ SCFT lies in the same supermultiplet as the energy-momentum tensor. This can be used to show that  the conformal anomaly is simply determined in terms of   't Hooft anomalies of the \R-current  by the following well-known formulae for the central charges \cite{Anselmi:1997am,Intriligator:2003jj,Cassani:2013dba}: 
\begin{equation}\label{andc}
a = \frac{9}{32} \text{tr}R^3 - \frac{3}{32} \text{tr}R\;, \qquad c = \frac{9}{32} \text{tr}R^3 - \frac{5}{32} \text{tr}R\;.
\end{equation}
Using   \eqref{thooftn4} and \eqref{andc} it is then straightforward to compute the   central charges in $\mathcal{N}=4$ SYM 
\begin{equation}\label{ccN=4}
a_{\mathcal{N}=4} = c_{\mathcal{N}=4} = \frac{1}{4}d_{G}\;.
\end{equation}
%


\begin{table}[t]
\renewcommand{\arraystretch}{1.2}
\begin{center}
\begin{tabular}{@{\extracolsep{10 pt}} c c c c c c}
\toprule
Field   & $SO(2)_1$ & $SO(2)_2$  & $SO(2)_3$  & $U(1)_R^{\cN=4}$ & $U(1)_R^{\rm LS}$   \\
\midrule
$\lambda$  &  $\phantom{-}{1\over 2}$ & $\phantom{-}{1\over 2}$  &  $\phantom{-}{1\over 2}$   & $\phantom{-}1$ & $\phantom{-}1$ 
	\\
$\chi_1$ & $-{1\over 2}$ &  $-{1\over 2}$ &   $\phantom{-}{1\over 2}$ &  $-{1\over 3}$ & $\phantom{-}*$   \\
$\chi_2$ &  $-{1\over 2}$ &  $\phantom{-}{1\over 2}$ &   $-{1\over 2}$ &  $-{1\over 3}$  &   $-{1\over 2}$   \\
$\chi_3$ &  $\phantom{-}{1\over 2}$ &  $-{1\over 2}$ &   $-{1\over 2}$ &  $-{1\over 3}$ &   $-{1\over 2}$   \\
\noalign{\smallskip}
\bottomrule
\end{tabular}
\caption{\label{table:one}
 Charges of the fermions in $\mathcal{N}=4$ SYM under various Abelian subgroups of $SO(6)$ discussed in the text.}
\end{center}
\end{table}

The LS fixed point \cite{Leigh:1995ep} is a strongly coupled $\mathcal{N}=1$ SCFT that can be thought of as the IR fixed point of an RG flow obtained by deforming $\mathcal{N}=4$ SYM with a particular relevant deformation of the superpotential
\begin{equation}\label{LSsuperpot}
\Delta W =\frac{m}{2}\,\Tr \Phi_1^2\;.
\end{equation}
Under the RG flow, the  superfield $\Phi_1$  is integrated out and the IR fixed point theory has the superconformal \R-symmetry given by\footnote{For notational clarity, from now on we will usually drop the superscript LS.}
\begin{equation}\label{TRLS}
T_R^{\rm LS} = \frac{1}{2}\left(T_1+T_2+2\,T_3\right)\;.
\end{equation}
The charges of the three remaining fermions under this \R-symmetry  are given in Table~\ref{table:one}. The charges of the two complex scalars, $\phi_{2}$ and $\phi_3$,  in the two remaining adjoint chiral multiplets of the LS fixed point under $U(1)_R$ are the same and equal to  $1/2$. In addition, there is an
 $SU(2)_F$ flavor symmetry which acts on the two chiral superfields $\Phi_{2}$ and $\Phi_3$ \cite{Leigh:1995ep}. This  $SU(2)_F$ is the $SU(2)_{\ell}$ factor in the decomposition $SU(2)_{\ell}\times SU(2)_r\times SO(2)\simeq SO(4)\times SO(2) \subset SO(6)$.  
 
Similarly as in the $\cN=4$ theory, one can calculate the cubic and linear 't Hooft anomalies for $U(1)_R$:
\begin{equation}
\mathop\text{tr}R^3_{\rm LS} = d_{G} [1^3-2(1/2)^3] = \frac{3}{4}d_G\;, \qquad \mathop\text{tr}R_{\rm LS} = d_{G} [1-2(1/2)] = 0\;,
\end{equation}
and find that the  central charges of the LS fixed point are 
\begin{equation}\label{ccLS}
a_{\rm LS} = c_{\rm LS} = \frac{27}{128}d_{G}\;.
\end{equation}
This yields  the familiar result (see \cite{Freedman:1999gp,Tachikawa:2009tt})
\begin{equation}\label{ccLS2}
\frac{a_{\rm LS}}{a_{\mathcal{N}=4}} = \frac{c_{\rm LS}}{c_{\mathcal{N}=4}}= \frac{27}{32}\;.
\end{equation}
%

\subsection{${(0,2)}$ SCFTs from the LS fixed point}
\label{subsec:LSFT}

One way to preserve some supersymmetry when a supersymmetric QFT is put on a curved manifold is to embed the structure group of the manifold into the $R$-symmetry of the QFT \cite{Witten:1988ze}. In other words, one has to turn on an $R$-symmetry background gauge field which cancels the spin-connection of the curved manifold. 
If the supersymmetric QFT at hand has additional flavor symmetry, one is also free to turn on a background gauge field for this flavor symmetry without any additional breaking of supersymmetry.

To construct the two-dimensional SCFTs of interest, we place the four-dimensional supersymmetric LS theory on $\mathbb{R}^{1,1}\times \Sigma_{\frak{g}}$ and perform a partial topological twist by turning on a background flux for the global symmetries. After integrating out all massive KK modes on $\Sigma_{\mathfrak{g}}$ we are left with an effective two-dimensional theory with $(0,2)$ supersymmetry. This is an old idea first explored in the current context in \cite{Bershadsky:1995vm,Maldacena:2000mw} and subsequently generalized in many papers including \cite{Benini:2013cda,Benini:2012cz,Almuhairi:2011ws,Naka:2002jz}.  Based on these results it is natural to assume that the low-energy theory will be a $(0,2)$ SCFT. The consistency of the results below together with the holographic analysis in the subsequent sections provide strong evidence for the validity of this assumption.

The four Poincar\'e supercharges of $\mathcal{N}=4$ SYM are in the $(\bf{2}, \bf{4})$ representation of $SO(1,3)\times SO(6)$. After the relevant deformation in \eqref{LSsuperpot} is turned on, only one of these supercharges is preserved and this is the supercharge present at the $\mathcal{N}=1$ LS   fixed point.\footnote{Since the LS theory is superconformal, there is also the corresponding conformal supercharge.} The supercharge has the same charges under $U(1)_R$ and $U(1)_F$ as the gaugino $\lambda$ in Table \ref{table:two}. After putting the four-dimensional theory on $\mathbb{R}^{1,1}\times \Sigma_{\frak{g}}$, in order to implement the topological twist and preserve some supersymmetry, we turn on a background gauge field along the generator
\begin{equation}\label{Tback}
T_{\mathfrak{b}} =\frac{\kappa}{2}\, T_R^{\rm LS} +\mathfrak{b} \,T_F \;,
\end{equation}
where 
\begin{equation}\label{TF}
T_{F} = \frac{1}{2}(T_1-T_2)\;,
\end{equation}
is the Cartan generator of the $SU(2)_F$ flavor symmetry.
The constant, $\kappa$, in  \eqref{Tback} is the normalized curvature of the Riemann surface, with $\kappa=1$ for $\mathfrak{g}=0$, $\kappa=0$ for $\mathfrak{g}=1$, $\kappa=-1$ for $\mathfrak{g}>1$. The coefficient $\mathfrak{b}$ is real and, since it measures the flux through a compact Riemann surface, it is quantized as $2(\mathfrak{g}-1)\mathfrak{b} \in \mathbb{Z}$ for $\mathfrak{g}\neq 1$ and $\mathfrak{b} \in \mathbb{Z}$ for $\mathfrak{g}= 1$. 


\begin{table}[t]
\renewcommand{\arraystretch}{1.2}
\begin{center}
\begin{tabular}{@{\extracolsep{10 pt}} c c c r r}
\toprule
Field  & $U(1)_R^{\rm LS}$ & $U(1)_F$ & $U(1)_\mathfrak{b}$ &   $U(1)_{\text{tr}}$ \\
\midrule
$\lambda$ & $\phantom{-}1$ & $\phantom{-}0$ & ${\kappa\over 2}$ & $1$ \\
$\chi_2$  &  $-{1\over 2}$  &   $-{1\over 2}$  &   $-{\kappa\over 4}-{\mathfrak{b}\over 2}$ & $-{1\over 2}-{\epsilon\over 2}$  \\
$\chi_3$ &  $-{1\over 2}$ &   $\phantom{-}{1\over 2}$  &   $-{\kappa\over 4}+{\mathfrak{b}\over 2}$ & $-{1\over 2}+{\epsilon\over 2}$ \\
\noalign{\smallskip}
\bottomrule
\end{tabular}
\caption{\label{table:two}
 Charges of the fermionic fields of the LS fixed point under the Abelian subgroups of $SU(2)_F\times U(1)_R^{\rm LS}$ discussed in the text.}
\end{center}
\end{table}

An analysis similar to the one in Appendix E of \cite{Benini:2013cda} shows that for $\mathfrak{g}\neq 1$ and any value of $\frak{b}$ there is $(0,2)$ supersymmetry preserved in the two-dimensional theory. When $\mathfrak{g}=1$ the discussion is slightly different. For $\mathfrak{g}=1$ and $\frak{b}=0$ there is $(2,2)$ supersymmetry preserved in two dimensions since the torus is flat. However, for $\mathfrak{g}=1$ and $\frak{b}\neq 0$, only two supercharges are preserved and one has $(0,2)$ supersymmetry \cite{Almuhairi:2011ws,Benini:2013cda,Kutasov:2013ffl}. Finally, one can preserve more supersymmetry in two-dimensions if the relevant deformation in \eqref{LSsuperpot} is switched off and one is left with a topological twist of the $\mathcal{N}=4$ theory. This was explored in detail in \cite{Benini:2013cda,Benini:2012cz}.

 We assume that in the   IR the effective two-dimensional theory is conformal.   Since it preserves $(0,2)$ supersymmetry, we can leverage anomalies to calculate the left and right central charges. The calculation proceeds as in \cite{Benini:2013cda}. The two-dimensional theory has two Abelian global symmetries: the $R$-symmetry of the LS fixed point given by the generator   \eqref{TRLS} as well as the $U(1)_F$  subgroup of the $SU(2)_F$ flavor symmetry with the generator   \eqref{TF}. The charges of the fermions under these two Abelian symmetries are given in Table \ref{table:two}. The background gauge field is along the generator \eqref{Tback} and has flux proportional to the volume form of the Riemann surface.

The two-dimensional superconformal $R$-symmetry is some linear combination of the Abelian global symmetries of the four-dimensional LS theory 
\begin{equation}\label{Ttr}
T_{\rm tr} = \epsilon~ T_F + T_R^{}\;.
\end{equation}
The real parameter $\epsilon$ is yet undetermined and will be fixed by applying the $c$-extremization procedure \cite{Benini:2012cz,Benini:2013cda}.\footnote{We assume that there are  no accidental Abelian symmetries emerging at the IR fixed point. } To this end we have to calculate the right-moving trial central charge, $c_r^{\rm tr}(\epsilon)$. 

The global symmetries of the two-dimensional theory are anomalous since there are massless chiral fermions. The number of these fermions is computed by the index theorem as in \cite{Maldacena:2000mw,Benini:2012cz,Benini:2013cda}
\begin{equation}\label{nrnldef}
n_r^{(\sigma)}-n_{\ell}^{(\sigma)} = -t_{\frak{b}}^{(\sigma)}\eta_{\Sigma}\;, \qquad \sigma=\lambda, \chi_2, \chi_3 \;,
\end{equation}
where $\eta_{\Sigma}=2|\mathfrak{g}-1|$ for $\mathfrak{g}\neq1$, $\eta_{\Sigma} =1$ for $\mathfrak{g}=1$, and $t_{\frak b}^{(\sigma)}$ is the charge of each of the three species of four-dimensional fermion fields under the background gauge field \eqref{Tback}. The values of $t_{\frak{b}}^{(\sigma)}$ are given in Table \ref{table:two}.

The two-dimensional (right-moving) trial central charge is computed using the fact that in two-dimensional $(0,2)$ SCFTs the conformal anomaly is proportional to the quadratic 't Hooft anomaly of the unique superconformal $U(1)$ $R$-symmetry \cite{AlvarezGaume:1983ig}
\begin{equation}\label{ctr1}\begin{split}
c_{r}^{\text{tr}} &= 3 \,d_{G}\ds\sum_{\sigma}(n_r^{(\sigma)}-n_{\ell}^{(\sigma)}) (q_{\rm tr}^{(\sigma)})^2\\ & = -3 \,\eta_{\Sigma}\, d_{G} \left[ t_{\frak{b}}^{(\lambda)}(q_{\rm tr}^{(\lambda)})^2 +t_{\frak{b}}^{(\chi_2)}(q_{\rm tr}^{(\chi_2)})^2 + t_{\frak{b}}^{(\chi_3)}(q_{\rm tr}^{(\chi_3)})^2 \right]\;,
\end{split}
\end{equation}
where $d_{G}$ is the dimension of the gauge group and $q_{\rm tr}^{(\sigma)}$ are the charges of the four-dimensional fermions under the trial $R$-symmetry \eqref{Ttr}  given in Table \ref{table:two}. Plugging these charges in \eqref{ctr1}, we find
\begin{equation}\label{ctr2}
c_{r}^{\text{tr}}(\epsilon,\frak b) = \frac{3}{8}\, \eta_{\Sigma}\, d_{G}\,(3-\epsilon^2+4\,\mathfrak{b}\,\epsilon)\;.
\end{equation}
Next, we extremize $c_{r}^{\text{tr}}(\epsilon,\frak b)$ with respect to $\epsilon$, which gives
\begin{equation}
\epsilon=2\mathfrak{b}\;,
\end{equation}
so that the two-dimensional right-moving central charge  is
\begin{equation}\label{cc2db}
c_{r}(\frak b) = \frac{3}{8}\,\eta_{\Sigma}\,d_G\, (3+ 4 \mathfrak{b}^2)\;.
\end{equation}

The central charge in \eqref{cc2db} is always positive and this suggests that, for fixed $G$ and $\mathfrak{g}$, there is a one-parameter family of two-dimensional $(0,2)$ SCFTs obtained by compactifying the LS fixed point on $\Sigma_{\mathfrak{g}}$ and turning on a flavor flux for the $U(1)_F$ flavor symmetry with magnitude $\mathfrak{b}$. For $\mathfrak{b}=0$ there is no flavor flux and one has
\begin{equation}\label{c2db=0}
c_{r}({ \mathfrak{b}=0}) = \frac{9}{8}\,\eta_{\Sigma}\,d_G = \frac{16}{3}\,\eta_{\Sigma} \,a_{\rm LS}\;.
\end{equation}
This is the universal two-dimensional $(0,2)$ fixed point that one can find for any four-dimensional $\mathcal{N}=1$ SCFT compactified on  $\Sigma_{\mathfrak{g}}$ \cite{Benini:2013cda,BBC}. Since the flavor flux vanishes, this theory   has an enhanced global symmetry given by $U(1)_R\times SU(2)_F$. For general values of the flavor flux $\mathfrak{b}$, the two-dimensional IR SCFTs have only $U(1)_R\times U(1)_F$ global symmetry.

One can also show that the family of two-dimensional SCFTs does not have a gravitational anomaly by evaluating the difference between the left and right central charges \cite{AlvarezGaume:1983ig}
\begin{equation}\label{crmincl}
c_{r}-c_{l} = d_{G}\ds\sum_{\sigma}\big(n_r^{(\sigma)}-n_{\ell}^{(\sigma)}\big) = 0\;.
\end{equation}
This result can be traced back to the fact that the LS fixed point has no linear 't Hooft anomaly for the superconformal $R$-symmetry, i.e. $a_{\rm{LS}}=c_{\rm{LS}}$ in \eqref{ccLS}.

The central charge, $c_r(\frak b)$, given by   \eqref{cc2db} is positive for any value of the flux parameter, $\mathfrak{b}$, and the genus of the Riemann surface, $\mathfrak{g}$. This means that unitarity is not violated and naively suggests that for any choice of these parameters there is a two-dimensional CFT in the IR.  We will discuss this further in Section \ref{Sec:AdS3sol}.

\subsection{$(0,2)$ SCFTs from $\mathcal{N}=4$ SYM}
\label{subsec:N=4flowsFT}

There is a natural generalization of the foregoing discussion, which suggests that the  $(0,2)$ SCFTs in Section \ref{subsec:LSFT} can also be accessed by a family of  RG flows in $\cN=4$ SYM. The idea is to turn on simultaneously two relevant deformations of $\mathcal{N}=4$ SYM:  the mass deformation \eqref{LSsuperpot} and  a twisted compactification along $\Sigma_\fg$.

Since the mass deformation \eqref{LSsuperpot} breaks the $R$-symmetry of $\mathcal{N}=4$ SYM from $SO(6)$ to $U(1)_R\times SU(2)_F$ and preserves $\cN=1$ supersymmetry, a simultaneous partial topological twist, with the background gauge field in this unbroken global symmetry subgroup,  will result in   a two-dimensional $(0,2)$ supersymmetric theory in the IR.  Assuming that the theory is superconformal, one can proceed as in Section \ref{subsec:LSFT} to calculate its central charge. We parametrize the background flux   as in $\eqref{Tback}$ and observe  that, with no  other  Abelian global symmetries available, 
 the trial $R$-symmetry is the same as in \eqref{Ttr}. We can also use  the formula \eqref{nrnldef} for the zero modes, except that we must now include  the fermions $\chi_1$ in the calculation. This leads to the following  trial central charge
\begin{equation}\label{ctr1N4}
\begin{split}
c_{r}^{\text{tr}} &= -3 \,\eta_{\Sigma} \,d_{G} \left[ t_{\frak{b}}^{(\lambda)}(q_{\rm tr}^{(\lambda)})^2 +t_{\frak{b}}^{(\chi_1)}(q_{\rm tr}^{(\chi_1)})^2+t_{\frak{b}}^{(\chi_2)} (q_{\rm tr}^{(\chi_2)})^2 + t_{\frak{b}}^{(\chi_3)}(q_{\rm tr}^{(\chi_3)})^2 \right] \\[6 pt]
 &= \frac{3}{8} \,\eta_{\Sigma} \,d_{G}\,(3-\epsilon^2+4\mathfrak{b}\epsilon)\;.
\end{split}
\end{equation}
This expression for $c_{r}^{\text{tr}}$ is the same as     \eqref{ctr2} for the simple reason that   the fermions $\chi_1$ have vanishing charges under $T_F$ and $T_R $. Then    $t_{\frak{b}}^{(\chi_1)}=0$ and thus   \eqref{ctr1N4} reduces  to \eqref{ctr1}. One can now extremize $c_{r}^{\text{tr}}$ as a function of $\epsilon$ to find that the  right-moving central charge is given by \eqref{cc2db} and the resulting $(0,2)$ SCFTs are the same as in Section \ref{subsec:LSFT}.

However, there is also a new feature of these RG flows   not present in Section \ref{subsec:LSFT}. 
For a fixed value of the flavor flux $\mathfrak{b}$, the trajectory of the RG flow 
 from $\mathcal{N}=4$ SYM  may not be unique given that  now one has two scales with which to deform, namely, the mass $m$ and the volume of the Riemann surface. Thus,  for a given value of $\frak b$, there should be a one parameter family of RG flow trajectories which connect $\mathcal{N}=4$ SYM with the corresponding two-dimensional $(0,2)$ SCFT. In contrast, the RG flow trajectory from the LS fixed point to the same  SCFT should be unique. We will see how this expectation bears out in Section \ref{Sec:HoloRG}, where we construct explicitly the holographic duals of those RG flows in gauged supergravity.

We would like to emphasize that the RG flows from $\mathcal{N}=4$ SYM obtained by a twisted compactification and a simultaneous mass deformation   are different from the RG flows studied in \cite{Benini:2012cz,Benini:2013cda} where  the $\mathcal{N}=4$ theory is deformed \emph{only} by a twisted compactification with no mass deformation. Then the effective two-dimensional $(0,2)$ SCFT has a $U(1)^3$ global symmetry (the Cartan of $SO(6)$) and in the  $c$-extremization calculation one must take an arbitrary linear combination of \emph{all} three $U(1)$ symmetries. In contrast, in the presence of the mass deformation  \eqref{LSsuperpot},  we have only $U(1)^2$ global symmetry (the Cartan of $U(1)_R\times SU(2)_F $) in the effective two-dimensional theory. This modifies the $c$-extremization calculation and the resulting central charge. The difference between the two deformations is most clearly visible by simply plugging the values of the background fluxes given in \eqref{Tback} into the formula for the   central charge  in equation (3.12) of \cite{Benini:2013cda}. This does not reproduce the  correct result  in \eqref{cc2db} above.

\sect{The supergravity model and BPS equations}
\label{Sec:SUGRAtrunc}

Our goal now is to find a dual gravity description of the SCFTs and RG flows discussed in Section~\ref{Sec:FieldTheory}. In this section we identify  a suitable truncation  of $\cN=8$, $d=5$ gauged supergravity  \cite{PPvN,Gunaydin:1984qu,Gunaydin:1985cu} and derive the corresponding BPS equations.

\subsection{The truncation}

By the AdS/CFT correspondence, the $SO(6)$ gauge symmetry of $\cN=8$, $d=5$ gauged supergravity can be identified with the $R$-symmetry of $\cN=4$ SYM. We are interested in a truncation of this theory which is invariant under  $U(1)_R\times U(1)_F\subset SO(6)$, where $U(1)_R$ is the $R$-symmetry  \eqref{TRLS}  of the LS fixed point and $U(1)_F\subset SU(2)_\ell$ is the flavor symmetry \eqref{TF}. The bosonic fields of $\cN=8$, $d=5$ gauged supergravity invariant under this subgroup of $SO(6)$ are:  the metric, $g_{\mu\nu}$, three vector fields, $A^{(i)}$, $i=1,2,3$, gauging the $SO(2)_1\times SO(2)_2\times SO(2)_3$ subgroup in $ SO(6)$, and six scalar fields parametrizing the coset
\begin{equation}\label{slcosetM}
\mathcal{M} = O(1,1)\times O(1,1) \times  \dfrac{SU(2,1)}{SU(2)\times U(1)}\,.
\end{equation}
Details of this truncation are discussed in  Appendix~\ref{appendixA}. Here let us note that the scalar fields, $\alpha$ and $\beta$, parametrizing the first two factors in   \eqref{slcosetM}, come from the scalars in $\bf{20}'$ of $SO(6)$ and are dual to the bosonic bilinear operators 
\begin{equation}\label{N=4ops}
\begin{split}
\mathcal{O}_{\alpha} &= {\rm Tr}\big(2\,|\phi_1|^2-|\phi_2|^2 - |\phi_3|^2\big)\;, \\[6 pt]
\mathcal{O}_{\beta} &= {\rm Tr}\big(\,|\phi_2|^2 - |\phi_3|^2\,\big)\;, 
\end{split}
\end{equation}
where $\phi_a$ are the three complex adjoint scalars of $\mathcal{N}=4$ SYM.\footnote{One should recall that the operator ${\rm Tr}\big(|\phi_1|^2+|\phi_2|^2 +|\phi_3|^2\big)$ receives a large anomalous dimension at strong 't Hooft coupling and is thus not a supergravity mode \cite{Witten:1998qj}. It can be added to \eqref{N=4ops} to preserve supersymmetry and positivity.}  
The last factor in \eqref{slcosetM} is parametrized by the complex scalar  $\chi e^{i\theta}$  in the  $\bf 10\oplus\overline{10}$ of $SO(6)$, plus the dilaton and axion dual to the complexified gauge coupling of $\mathcal{N}=4$ SYM. For  the solutions we are interested in, one can consistently turn off the dilaton and axion and set  the phase $\theta$ to be constant. This leaves only one real scalar, $\chi$, in the third factor in \eqref{slcosetM}, which is dual to the fermion bilinear 
\begin{equation}\label{Ochi}
\mathcal{O}_{\chi} = {\rm Tr} (\chi_1\chi_1 + {\rm h.c.})\;,
\end{equation}
where, as in Section \ref{subsec:ANSYM}, $\chi_a$ are the adjoint Weyl fermions in the three chiral super fields. 
  
The bosonic part of the action in this sector is derived in Appendix~\ref{appendixA}. For the  trivial dilaton/axion, it reads\footnote{We follow the same conventions as in   \cite{Gunaydin:1985cu,Bobev:2010de} with  the mostly minus  metric.  This action can also be obtained  from  another truncation of the $\cN=8$ supergravity discussed in  \cite{Bobev:2010de,Khavaev:2000gb}. In the notation of those papers we are keeping the fields $\alpha$, $\beta$, $\theta_1\equiv\theta$ and $\varphi_1\equiv\chi$.}
\begin{equation}\label{CCaction}
\begin{split}
e^{-1}\cals L&  \eql  -{1\over 4} R -{1\over 4}\Big[e^{4(\alpha-\beta)}\,F^{(1)}_{\mu\nu}F^{(1)}{}^{\mu\nu}+e^{4(\alpha+\beta)} F^{(2)}_{\mu\nu}F^{(2)}{}^{\mu\nu}+e^{-8\alpha}F^{(3)}_{\mu\nu}F^{(3)}{}^{\mu\nu}\Big]
\\[5pt] & +3(\partial_\mu\alpha)^2+(\partial_\mu\beta)^2+{1\over 2}(\partial_\mu\chi)^2+{1\over 8}\sinh^2(2\chi)\Big[\partial_\mu\theta+{g}\,(A^{(1)}+A^{(2)}-A^{(3)})\Big]^2 -g^2\,\cals P\,,
\end{split}
\end{equation}
with the scalar potential 
\begin{equation}\label{CCpotential}
\begin{split}
\cals P\eql {1\over 8}e^{-4(\alpha+\beta)}\cosh^2\chi\Big[
\big(1+e^{8\beta}+e^{4(3\alpha+\beta)}\big)\cosh^2\chi-\big(1+e^{4\beta}+e^{2(3\alpha+\beta)}\big)^2
\Big]\,.
\end{split}
\end{equation}
As in similar supersymmetric truncations (see, e.g. \cite{Freedman:1999gp,Bobev:2010de}),  the potential can be  rewritten as
\begin{equation}
\mathcal{P} =\frac{1}{48} \left(\partial_{\alpha}\mathcal{W}\right)^2 + \frac{1}{16} \left(\partial_{\beta}\mathcal{W}\right)^2 + \frac{1}{8} \left(\partial_{\chi}\mathcal{W}\right)^2- \frac{1}{3}\mathcal{W}^2\;,
\end{equation}
where  
\begin{equation}\label{suppot5d}
\cals  W\eql \frac{1}{4} e^{-2 (\alpha +\beta )} \left(e^{6 \alpha +2 \beta } (\cosh (2 \chi )-3)-2 \left(e^{4 \beta
   }+1\right) \cosh ^2(\chi )\right)\,,
\end{equation}
is the superpotential determined by the supersymmetry variations, see Section~\ref{BPSeqs}. 

The potential, $\cP$,  has three critical points \cite{Khavaev:1998fb} and those give rise to three $AdS_5$ vacua in this truncation:
\begin{itemize}
\item [(i)] The maximally supersymmetric critical point with $SO(6)$ global symmetry:
\begin{equation}\label{SO6fp}
\alpha =\beta=\chi=0\;, \qquad \mathcal{P} = -\frac{3}{4}\;, \qquad L = \frac{2}{g}\;.
\end{equation}
\item [(ii)] The $\mathcal{N}=2$ critical point with $SU(2) \times U(1)$ global symmetry:
\begin{equation}\label{PWfp}
\alpha = \frac{1}{6}\log 2\;, \qquad \beta=0\;, \qquad \chi=\pm \frac{1}{2}\log 3\;, \qquad \mathcal{P} = -\frac{2^{4/3}}{3}\;, \qquad L = \frac{3}{2^{2/3}g}\;.
\end{equation}

\item [(iii)] The non-supersymmetric $SU(3)$-invariant critical point:
\begin{equation}\label{SU3fp}
\alpha =0\;, \qquad \beta=0\;, \qquad \chi= \frac{1}{2}\log (2\pm \sqrt{3})\;, \qquad \mathcal{P} = -\frac{27}{32}\;, \qquad L = \frac{2^{5/2}}{3g}\;.
\end{equation}
\end{itemize}
The $AdS_5$ radius, $L$, is related to the critical value, $\cals P_*$, of the potential by
\begin{equation}\label{}
L^2\eql -{3\over g^2\,\cals P_*}\,.
\end{equation}
Both supersymmetric  points (i) and (ii) are also critical points of the superpotential \eqref{suppot5d} and are perturbatively stable. The  $SU(3)$-invariant critical point is stable within the truncation, but is perturbatively unstable in the full $\cN=8$ theory.

In the following we will concentrate on the 
 supersymmetric $SU(2) \times U(1)$-invariant critical point (ii),   from now on referred to as the KPW point, which is  the holographic dual of the $\mathcal{N}=1$  LS SCFT \cite{Freedman:1999gp,Khavaev:1998fb,Karch:1999pv}. In particular,  within the truncation    \eqref{CCaction}, we will be interested in constructing supersymmetric  flows that, in a certain sense,\footnote{See, Section~\ref{Sec:HoloRG} for a detailed discussion.}  interpolate between this point and  supersymmetric solutions of the form $AdS_3\times \Sigma_\mathfrak{g}$. 

It might be worth pointing out that the five-dimensional theory used in \cite{Maldacena:2000mw,Benini:2012cz,Benini:2013cda} to find similar supersymmetric flows from the maximally supersymmetric $SO(6)$ point  to $AdS_3\times \Sigma_\mathfrak{g}$ solutions  is a truncation of  \eqref{CCaction} obtained by setting $\chi=\theta=0$. This truncation is usually  called  the STU model of five-dimensional gauged supergravity \cite{Cvetic:1999xp}.

\subsection{The Ansatz}

We will assume from now on  that the metric on the Riemann surface, $\Sigma_{\frak g}$, has the constant curvature. This is justified if we extrapolate  the result in \cite{Anderson:2011cz}, where it was shown that  the holographic RG flow uniformizes the metric on a Riemann surface for the supergravity truncation with $\chi=0$. We have checked that for  $\mathfrak{g}=0,1$  there are no   supersymmetric $AdS_3\times \Sigma_\mathfrak{g}$ solutions and, to simplify the presentation, we will assume from now on that $\frak g>1$. The Riemann surface $\Sigma_{\mathfrak{g}}$ can then be represented as a quotient of the upper half plane with the metric,
\begin{equation}\label{}
ds^2_{\mathbb{H}_2}\eql {1\over y^2}\left({dx^2+dy^2}\right)\,,
\end{equation}
by a discrete subgroup of $PSL(2,\RR)$. This makes our choice of a constant curvature metric manifest. 

To find the supergravity solutions dual to the SCFTs in Section~\ref{Sec:FieldTheory}, we employ the same   Ansatz as in  \cite{Maldacena:2000mw,Benini:2012cz,Benini:2013cda}, where the metric is of the form 
\begin{equation}\label{}
ds^2 = e^{2f(r)}(dt^2-dz^2-dr^2)-{e^{2h(r)}\over y^2}(dx^2+dy^2)\,,
\end{equation}
with two undetermined functions, $f(r)$ and $h(r)$, of the radial coordinate, $r$. This metric Ansatz encompasses two types of solutions we are interested in:  (i)   asymptotically locally $AdS_5$ solutions\footnote{For a formal definition and a review of asymptotically locally $AdS$ spacetimes in holography, see \cite{Skenderis:2002wp}.} with $\mathbb{R}^{1,1}\times \Sigma_{\frak{g}}$ boundary and  $f(r)\sim h(r)\sim-\log r$ diverging at the same rate for $r\rightarrow 0$, and  (ii)~$AdS_3\times \Sigma_\mathfrak{g}$ solutions with constant $h(r)$ and divergent $f(r)\sim -\log r$ for $r\rightarrow\infty $.
When needed, we adopt the obvious choice of frames
\begin{equation}\label{framedef}
e^{0} = e^{f(r)}dt\;, \qquad e^{1} = e^{f(r)}dz\;, \qquad e^{2} = e^{f(r)}dr\;, \qquad e^{3} = \frac{e^{h(r)}}{y}dx\;, \qquad e^{4} = \frac{e^{h(r)}}{y}dy\;.
\end{equation}

The topological twist in the dual field theory  implies that the   flux of the gauge field on the gravity side  must be  proportional to the volume of the Riemann surface,
\begin{equation}\label{theflux}
F^{(i)}~\equiv~ dA^{(i)}\eql a_i\,{\rm vol}_{\Sigma_{\mathfrak g}}\,,\qquad {\rm vol}_{\Sigma_{\mathfrak g}}\eql {1\over y^2}\,dx\wedge dy \,,\qquad i=1,2,3\,,
\end{equation}
where $a_i$ are arbitrary constants. Correspondingly, we take the gauge field potentials to be
\begin{equation}\label{}
A^{(i)}\eql {a_i\over y}\,dx\,,\qquad i=1,2,3\,.
\end{equation}
Finally, the scalar fields, $\alpha(r)$, $\beta(r)$, $\chi(r)$ and $\theta(r)$,  depend only on the radial coordinate. 

With this Ansatz at hand one can derive a system of BPS equations directly from the  supersymmetry variations of the  $\mathcal{N}=8$, $d=5$ supergravity. In the next section we outline the calculation and summarize the results.

\subsection{The BPS equations}
\label{BPSeqs}

The supersymmetry variations of $\cN=8$, $d=5$ gauged supergravity read:\footnote{For  definitions of the various tensors and further details, we refer the reader to \cite{Gunaydin:1985cu}.} 
\begin{align}\label{var32}
\delta\psi_{\mu\,a} & \eql D_\mu\epsilon_a-{1\over 6}\,g\,W_{ab}\gamma_\mu\epsilon^b-{1\over 6}\,H_{\nu\rho\,ab}(\gamma^{\nu\rho}\gamma_\mu+2\gamma^\nu\delta^\rho{}_\mu)\, \epsilon^b\,,\\[6 pt]
\delta\chi_{abc} & \eql \sqrt 2\Big[\gamma^\mu P_{\mu\,abcd}\,\epsilon^d-{1\over 2}\,g A_{dabc}\,\epsilon^d-{3\over 4}\,\gamma^{\mu\nu} H_{\mu\nu[ab}\,\epsilon_{c]|}\Big]\,.
\label{var12}\end{align}
Under  $U(1)_R\times U(1)_F$, 
the eight gravitini, $\psi^a$, and the supersymmetry parameters, $\epsilon^a$, transform with  the charges
\begin{equation}\label{branch8}
\bfs 8\quad \longrightarrow \quad (0,0)+(0,0)+(1,0)+(-1,0)+(\coeff 1 2,\coeff 1 2)+(-\coeff 1 2,-\coeff 1 2)+(\coeff 1 2,-\coeff 1 2)+(-\coeff 1 2, \coeff 1 2)\,.
\end{equation}
In the following, we are interested in the sector where the gravitini and the corresponding  supersymmetries have the unit $R$-charge and are invariant under the flavor symmetry.  Those supersymmetry parameters are given by  (see (3.1) in \cite{Khavaev:2000gb} and  (3.36) in  \cite{Bobev:2010de}),
\begin{equation}\label{susypar}
\epsilon^a\eql \varepsilon_{(1)}\,\eta^a_{(1)}+\varepsilon_{(2)}\,\eta^a_{(2)}\,,\qquad \epsilon_a\eql \Omega_{ab}\,\epsilon^b\,,
\end{equation}
where $\varepsilon_{(1)}$ and $\varepsilon_{(2)}$ are a symplectic pair of five-dimensional spinors, $\Omega_{ab}$ is an $8\times 8$ symplectic matrix, and
\begin{equation}\label{}
\eta_{(1)}\eql (1,\, 0,-1,\, 0,\,  0,\,  1,\,  0,\,  1)
\,,\qquad 
\eta_{(2)}\eql  (0,\, 1,\, 0,\, 1,-1,\,  0,\,  1,\,  0)\,.
\end{equation}
In this two-dimensional subspace we have,
\begin{equation}\label{}
\Omega_{ab}\,\eta^b_{(i)}\eql \omega_{ij}\eta^a_{(j)}\,,\qquad 
W_{ab}\,\eta^b_{(i)}\eql \cals W\,\eta^a_{(i)}\,,
\end{equation}
where $\cals W$ is the superpotential \eqref{suppot5d} and we have defined $\omega_{12}=-\omega_{21}=1$, $\omega_{11}=\omega_{22}=0$.

We start with the spin-3/2 variations \eqref{var32} with $\epsilon^a$ in \eqref{susypar}. The $r$-dependence of the Killing spinors for the unbroken supersymmetries is determined by the vanishing of the spin-3/2 variation along the radial direction, 
\begin{equation}\label{varr}
\partial_r\varepsilon_{(i)}+{1\over 6}\,e^{f-2h} \,\cals H\,\gamma^{234}\varepsilon_{(i)}+\omega_{ij} 
\Big({g\over 6}\,\,e^f \,\cals W\, 
\gamma^2+{1\over 2}\,\sinh^2\chi\,\theta'
\Big)\varepsilon_{(j)}\eql 0\,,
\end{equation}
where the prime denotes a derivative with respect to $r$. Then, assuming that  the Killing spinors   do not depend on the $t$, $z$, $x$ and $y$ coordinates, the remaining variations reduce to  three algebraic constraints on $\varepsilon_{(1)}$ and $\varepsilon_{(2)}$: 
\begin{align}\label{vartx}
\big(3\,e^{-f}f'\,\gamma^2 - e^{-2h}\,\cals H\,\gamma^{34}\big)\,\varepsilon_{(i)} -{g}\,\omega_{ij}\,\cals W\, \varepsilon_{(j)} & \eql 0\,,\\[6 pt]
\label{varxy}
(3\,e^{-f}h'\,\gamma^2   +2\,e^{-2h}\,\cals H\,\gamma^{34}\big)\,\varepsilon_{(i)}-g\,\omega_{ij}\,\cals W\,\varepsilon_{(j)}& \eql 0\,,\\[6 pt]
\label{varxydef}
   2\, \gamma^4\,\varepsilon_{(i)}-{g }\,\omega_{ij}\, \, \Lambda\, \gamma^3\varepsilon_{(j)}& \eql 0\,.
\end{align}
Here and above, we have defined 
\begin{equation}\label{}
\begin{split}
\cals H  & \eql  e^{2\alpha-2\beta}a_1+e^{2\alpha+2\beta}a_2+e^{-4\alpha} a_3  \,,\\[6 pt]
\Lambda & \eql a_1+a_2+3 a_3+(a_1+a_2-a_3)\cosh 2\chi\,.
\end{split}
\end{equation}
The first equation, \eqref{vartx},  arises from the variations along $\mathbb{R}^{1,1}$, while the remaining two, \eqref{varxy} and \eqref{varxydef}, from the variations along the Riemann surface. The function $\cals H$ in  \eqref{varr}, \eqref{vartx} and \eqref{varxy}  is  the  eigenvalue of the $H_{ab}$ tensor in \eqref{var32} and its dependence on $a_i$'s comes from the field strengths, $F^{(i)}$. The $\Lambda$-term in \eqref{varxydef} comes from   the composite connection in the covariant derivative in \eqref{var32} and its dependence on $a_i$ is due to  the gauge potentials, $A^{(i)}$.

To solve the algebraic equations \eqref{vartx}-\eqref{varxydef} we impose  projection conditions on the Killing spinors,
\begin{equation}\label{projc}
\gamma^{34}\varepsilon_{(i)}\eql - \omega_{ij}\varepsilon_{(j)}\,,\qquad \gamma^2\varepsilon_{(i)}\eql \omega_{ij}\varepsilon_{(j)}\,,
\end{equation}
which are unique up to a choice of signs on the right hand side, with different choices leading to equivalent BPS equations. Note that by combining the two projectors in \eqref{projc} and using  
that $\gamma^0\gamma^1\gamma^2\gamma^3\gamma^4 = 1$ one finds 
\begin{equation}\label{}
\gamma^{01}\varepsilon_{(i)}\eql \varepsilon_{(i)}\,,
\end{equation}
which shows that  the two-dimensional holographic field theory indeed has $(0,2)$ supersymmetry.

Using \eqref{projc} in \eqref{varxydef} we then get
\begin{equation}\label{acond}
a_1+a_2+3 a_3+(a_1+a_2-a_3)\cosh 2\chi \eql {2\over g}\,.
\end{equation}
Since $a_i$ are constant, supersymmetric  flows with a varying field, $\chi$,  must satisfy\footnote{Note that for $\chi=0$, \eqref{acond} reduces to $a_1+a_2+a_3=1/g$, which leads to the solutions constructed in \cite{Benini:2013cda}.}
\begin{equation}\label{aconstr}
a_3=a_1+a_2\;, \qquad  a_1+a_2 = \dfrac{1}{2g}\;.
\end{equation}
It is convenient to solve the second constraint in \eqref{aconstr} by introducing a single parameter, $\mathfrak a$, %
\begin{equation}\label{a1a2a}
a_1 \eql  {1\over g}\Big({1\over 4}+\mathfrak a\Big)\,,  \qquad a_2 \eql  {1\over g}\Big({1\over 4}-\mathfrak a\Big)\,.
\end{equation}

Note that \eqref{acond}  is tantamount to the topological twist along the Riemann surface. Indeed, it implies a cancellation, in the covariant derivative  in \eqref{var32}, between the terms with the spin connection along the Riemann surface and the vector potential terms from the composite connection. This is also a supergravity manifestation of the fact that in order to preserve some supersymmetry we need to turn on a specific background gauge field for the $R$-symmetry. Indeed,  \eqref{theflux}, \eqref{aconstr} and \eqref{a1a2a} imply that
\begin{equation}\label{sugraR}
F^{(i)}\,T_i\eql \Big({1\over 2g}\,T_R+{2\mathfrak a\over g}\,T_F\Big)\,{\rm vol}_{\Sigma_{\mathfrak g}}\,,
\end{equation}
with a fixed component along $T_R$. This matches the field theory flux  \eqref{Tback} for $\kappa=-1$ provided  we identify $\frak{b}= - 2\frak{a}$ and set $g=2$.\footnote{The factor of 2 in $\mathfrak{b}=- 2\mathfrak{a}$ is due to a different normalization of the Maxwell terms in field theory and in supergravity. To see this, we compare the action  (46) in \cite{Maldacena:2000mw} with \eqref{CCaction} above. This leads to $a_i^{BB}=2a_i^{\rm here}$, where $a_i^{BB}$ are the constants used in \cite{Benini:2012cz,Benini:2013cda}. The sign difference between $\mathfrak{b}$ and $\mathfrak{a}$ comes from the opposite signature of the space-time metric   in field theory.} Using \eqref{theflux} we find that $a_i$ must be quantized  such that $4(\frak{g}-1)a_i \in \mathbb{Z}$. After using \eqref{aconstr}  and  \eqref{a1a2a} this is compatible with the quantization condition $2(\frak{g}-1)\frak{b} \in \mathbb{Z}$ discussed below \eqref{TF}.

Finally, using \eqref{projc} in \eqref{vartx} and \eqref{varxy}, we obtain  the following flow equations for the metric functions
\begin{equation}\label{BPSfh}
\begin{split}
f' & \eql {g\over 3}\Big( e^f \,\cals W-{1\over g }e^{f-2h}\,\cals H\Big)\,,\qquad 
h'  \eql {g\over 3}\Big( e^f \, \cals W+{2\over g }e^{f-2h}\,\cals H\Big)\,.
\end{split}
\end{equation}

The spin-1/2 variations \eqref{var12} simplify dramatically after using the projections \eqref{projc} and the vector field constraints \eqref{aconstr}. All variations  reduce to three first order flow equations for the three scalars,
\begin{equation}\label{BPSabc}
\begin{split}
\alpha' &\eql -{g\over 12}\,{\partial \over\partial\alpha}\Big( e^f\,  \cals W+{1\over g }e^{f-2h}\, \cals  H\Big)\,,\\[6 pt]
\beta'  & \eql -{g\over 4}\,{\partial \over\partial\beta}\Big( e^f\,  \cals W+{1\over g }e^{f-2h}\, \cals  H\Big)\,,\\[6 pt]
\chi'  & \eql -{g\over 2}\,{\partial \over\partial\chi}\Big( e^f\,  \cals W+{1\over g }e^{f-2h}\, \cals H\Big)\,,
\end{split}
\end{equation}
and set $\theta$ to be constant. Note that the BPS equations \eqref{BPSfh} and \eqref{BPSabc} are symmetric under $\frak{a}\rightarrow - \frak{a}$ and $\beta\rightarrow -\beta$.

The flow equations \eqref{BPSfh} and \eqref{BPSabc} allow for an explicit solution to \eqref{varr},
\begin{equation}\label{}
\varepsilon_{(i)}\eql e^{f/2}\,\varepsilon_{(i)}^0\,,\qquad i=1,2\,,
\end{equation}
where $\varepsilon_{(i)}^0$ are two constant spinors satisfying the same projections as in \eqref{projc}.

This completes our analysis of the supersymmetry variations \eqref{var32} and \eqref{var12}. One can check that the BPS equations \eqref{BPSfh} and \eqref{BPSabc} imply that the equations of motion are satisfied. In order to classify   supersymmetric $AdS_3\times \Sigma_{\mathfrak g}$ solutions in the next section, and to construct holographic RG flows in Section~\ref{Sec:HoloRG}, it will be sufficient to consider the first order ODEs  \eqref{BPSfh} and \eqref{BPSabc}.

\sect{$AdS_3\times \Sigma_{\frak{g}}$ solutions}
\label{Sec:AdS3sol}

To find $AdS_3\times \Sigma_{\frak{g}}$ solutions of the BPS equations   \eqref{BPSfh} and \eqref{BPSabc}, we take constant scalars and set 
\begin{equation}
f(r) =  f_0 + \log \frac{1}{r}\;,  \qquad h(r) = h_0\;,
\end{equation}
where $f_0$ and $h_0$ are constants. This turns \eqref{BPSfh} and \eqref{BPSabc} into algebraic equations that can be solved systematically. The result is  that all $AdS_3\times \Sigma_{\frak{}g}$ solutions of the BPS equations with $|\frak a|<1/4$ are given by
\begin{equation}\label{AdS3sol}
\begin{gathered}
e^{12\alpha}  = \frac{4}{1-16\mathfrak{a}^2}\;, \qquad  e^{4\beta} = \frac{1-4\mathfrak{a}}{1+4\mathfrak{a}}\;, \qquad e^{\chi} = \frac{2+\sqrt{1-16\mathfrak{a}^2}}{\sqrt{3+16\mathfrak{a}^2}}\;, \\[6 pt]
e^{3f_0}  = \frac{2}{g^3} (1-16\mathfrak{a}^2)\;, \qquad  e^{6h_0} = \frac{1}{16 g^6} \frac{(3+16\mathfrak{a}^2)^3}{1-16 \mathfrak{a}^2} \;.
\end{gathered}
\end{equation}

With the $AdS_3\times \Sigma_{\frak{g}}$ solutions at hand, we can now calculate the central charge in the dual field theory using the  Brown-Henneaux formula \cite{Brown:1986nw}, 
\begin{equation}\label{BHform}
c = \frac{3L^{(3)}}{2G_{N}^{(3)}}\;,
\end{equation}
where $L^{(3)}$ is the effective scale of $AdS_3$ and $G_{N}^{(3)}$ is the three-dimensional Newton constant. After setting $g=2$, so that we have the same normalization and conventions as in \cite{Benini:2012cz,Benini:2013cda,Maldacena:2000mw}, we find that the central charges of   field theories dual to the $AdS_3\times \Sigma_{\frak g}$ solutions \eqref{AdS3sol} are
\begin{equation}\label{ccsugra}
c = 6 \,\eta_{\Sigma }\,N^2 e^{f_0+2h_0} = \frac{6}{2 g^3}\, \eta_{\Sigma}\, N^2 (3+16 \mathfrak{a}^2) = \frac{3}{8}\, \eta_{\Sigma}\, N^2 (3+16 \mathfrak{a}^2)\;.
\end{equation}
Upon the identification $\mathfrak{b}=-2\mathfrak{a}$, this precisely reproduces the field theory result \eqref{cc2db} for $G=SU(N)$ and $N \gg 1$. 

The solution with $\mathfrak{a}=0$ is somewhat special. From \eqref{AdS3sol} we see that it has $\beta=0$ and, since $a_1-a_2$ vanishes, there is no flux for the gauge field along the generator $T_F$ in \eqref{TF}. This means that the $SU(2)$ gauge symmetry is not broken, which is in harmony with the fact that in field theory  we have an enhanced $SU(2)_F$ global symmetry precisely for $\mathfrak{b}=0$. As will be discussed in the next section,  the BPS equations with  $\mathfrak{a}=0$ admit an analytic solution which interpolates between the KPW point and the $AdS_3\times \Sigma_{\frak g}$ vacuum in \eqref{AdS3sol}.

When $\mathfrak{a}=\pm 1/4$, the supersymmetric $AdS_3\times \Sigma_{\frak{g}}$ solutions  \eqref{AdS3sol} cease to exist and one of the gauge fields $A^{(1)}$ or $A^{(2)}$ vanishes, see \eqref{theflux} and \eqref{a1a2a}. This happens because in solving the  BPS equations we have assumed that $\chi \neq 0$. For $\chi=0$,  there is a special family of solutions to  \eqref{BPSfh},  \eqref{BPSabc} and  \eqref{acond} given by
\begin{equation}\label{AdS3a14}
\begin{gathered}
\frak{a} = \pm \dfrac{1}{4}\;, \qquad\qquad e^{2\beta} = \dfrac{1}{2}\left(\mp e^{6\alpha}+\sqrt{4+e^{12\alpha}}\right)\;,  \\[6 pt]
e^{2h_0} = \dfrac{e^{-2\alpha}}{8}\left(e^{6\alpha}+\sqrt{4+e^{12\alpha}}\right)\;, \qquad e^{f_0} = \dfrac{e^{2\alpha}}{2}\left(-e^{6\alpha}+\sqrt{4+e^{12\alpha}}\right)\;,
\end{gathered}
\end{equation}
and parametrized by $\alpha$.  This family is not new -- it was found in \cite{Maldacena:2000mw} and recently discussed further in \cite{Benini:2013cda}. It has enhanced $(2,2)$ supersymmetry and the scalar $\alpha$ is a free modulus indicating the existence of an exactly marginal deformation in the dual field theory. The central charge for these solutions is independent of the sign in $\frak{a}=\pm 1/4$ and the value of the scalar $\alpha$, 
\begin{equation}
c = 6\eta_{\Sigma} N^2 e^{f_0+2h_0} =  \frac{3}{2}\eta_{\Sigma} N^2 = 3(\frak{g}-1) N^2\;.
\end{equation}
This is always an integer multiple of 3 and is precisely the central charge of the $(2,2)$ solutions found in \cite{Maldacena:2000mw,Benini:2013cda}.

The central charges \eqref{cc2db} computed via anomalies and $c$-extremization are positive, and thus compatible with unitarity, for all possible values of the genus $\frak{g}$ and the flavor flux $\frak{b}$. Here, we see that the $AdS_3\times \Sigma_{\frak{g}}$  supergravity solutions are regular and causal only for $\frak{g}>1$ and $|\mathfrak{a}|\leq 1/4$, or, equivalently,  $|\mathfrak{b}|\leq 1/2$.  For values of the parameters outside this range, holography suggests that one of the following scenarios might be realized: (i) The IR theory is not conformal; (ii)~There is an IR SCFT, but it does not admit a gravity dual; (iii) There are accidental Abelian symmetries in the IR which render the use of $c$-extremization invalid; (iv) There is an IR SCFT with a non-normalizable vacuum state.\footnote{An example where this is realized is the $(4,4)$ two-dimensional sigma model on the Hitchin moduli space obtained by placing $\mathcal{N}=4$ SYM on a Riemann surface with a special partial topological twist \cite{Bershadsky:1995vm}. Since the Hitchin moduli space is a non-compact hyper-K\"ahler manifold, the SCFT does not have a normalizable vacuum state and thus there is no dual $AdS_3$ vacuum \cite{Maldacena:2000mw}. 
}
(v) 
 Finally, it is also possible that there are $AdS_3 \times \Sigma_{\frak{g}}$ solutions for other values of the parameters $\frak{g}$ and $\frak{b}$ that are not captured by the present supergravity truncation.

Starting with the action in \eqref{CCaction} and setting $\chi=0$, we can recover all supersymmetric $AdS_3\times \Sigma_{\frak g}$ solutions found in \cite{Benini:2013cda}. Thus one may wonder whether it is possible to realize a holographic RG flow that interpolates between some of the $AdS_3\times \Sigma_{\frak g}$ solutions in \cite{Benini:2013cda} and the solutions   \eqref{AdS3sol} in this paper. If such a flow exists  within our supergravity truncation, the magnetic flux of the gauge field, specified by the constants $a_i$ in \eqref{theflux}, should not change along the flow. The reason is that, as discussed below \eqref{sugraR}, the parameters $a_i$ are quantized and thus cannot change continuously as a function of the radial variable, $r$. One can then show that there are no values of the parameters $a_i$ (except for $\frak{a}=\pm 1/4$, see the discussion above) for which there is both an $AdS_3$ solution in the truncation of \cite{Benini:2013cda} and a solution of \eqref{BPSfh} and  \eqref{BPSabc}. This means that within  the supergravity truncation we are using there are no holographic RG flows interpolating between the $AdS_3\times \Sigma_{\frak g}$ solutions \eqref{AdS3sol}  and the ones in~\cite{Benini:2013cda}.

\sect{Holographic RG flows}
\label{Sec:HoloRG}

We are looking for domain wall  solutions to the BPS equations  \eqref{BPSfh} and \eqref{BPSabc}
that are holographically  dual to the RG flows discussed  in Section~\ref{Sec:FieldTheory}. In the UV ($r\rightarrow 0$), such solutions should  asymptote to one of the  $AdS_5$  solutions for either the maximally supersymmetric $SO(6)$ critical point \eqref{SO6fp} 
or the KPW critical point \eqref{PWfp}. 
More precisely, since the field theory lives on $\mathbb{R}^{1,1}\times \Sigma_{\mathfrak{g}}$, the five-dimensional space-time is only asymptotically locally $AdS_5$,\footnote{With some abuse of terminology, we will refer  to the five-dimensional asymptotically locally $AdS_5$ solutions as   $AdS_5$ solutions or simply  critical points.} namely,   its metric on the boundary has a non-zero curvature, which is cancelled by the non-zero background flux determined by  the gauge fields \eqref{theflux}. 
In the IR ($ r\rightarrow \infty$), the  solutions should  asymptote  to one of  the $AdS_3\times \Sigma_{\frak g}$  points \eqref{AdS3sol}. Similar supersymmetric flow solutions in five-dimensional gauged supergravity were constructed in \cite{Benini:2013cda,Maldacena:2000mw,Almuhairi:2011ws,Donos:2011pn,Naka:2002jz,Cucu:2003yk,Hristov:2013xza}.

The structure of the BPS equations becomes particularly simple after we rewrite them    in terms of  a   ``superpotential''
\begin{equation}\label{Vsuperpot}
\mathcal{V} = g\, e^{-2h}\,  \cals W+e^{-4h}\, \cals  H\;,
\end{equation}
and  a new radial variable 
\begin{equation}\label{rhodef}
\rho = f+2h \qquad \Longrightarrow\qquad \frac{d\rho}{dr} = e^{\rho}\, \mathcal{V}\,.
\end{equation}
Indeed, if we   define the canonically  normalized scalar fields
\begin{equation}\label{varphidef}
\varphial  \equiv 2\sqrt{3} \alpha\;, \qquad \varphibe \equiv 2 \beta\;, \qquad \varphichi  \equiv \sqrt{2} \chi\;, \qquad \varphih  \equiv \sqrt{6} h\;,
\end{equation}
then  \eqref{BPSfh} and \eqref{BPSabc} are  equivalent to \eqref{rhodef} plus  the following first order system of  flow equations
\begin{equation}\label{BPSeqnV}
\frac{d\varphi_i}{d\rho} = - {1\over \cals V}\,\frac{\partial\mathcal{V}}{\partial\varphi_i}\;, \qquad\qquad i=\alpha,\beta,\chi,h\,,
\end{equation}
where $\cals V$ is now a function of the fields, $\varphi_i$, but not $f$, and depends on the flux parameter, $\fa$.

%
\begin{figure}[t]
\centering
\includegraphics[width=5cm]{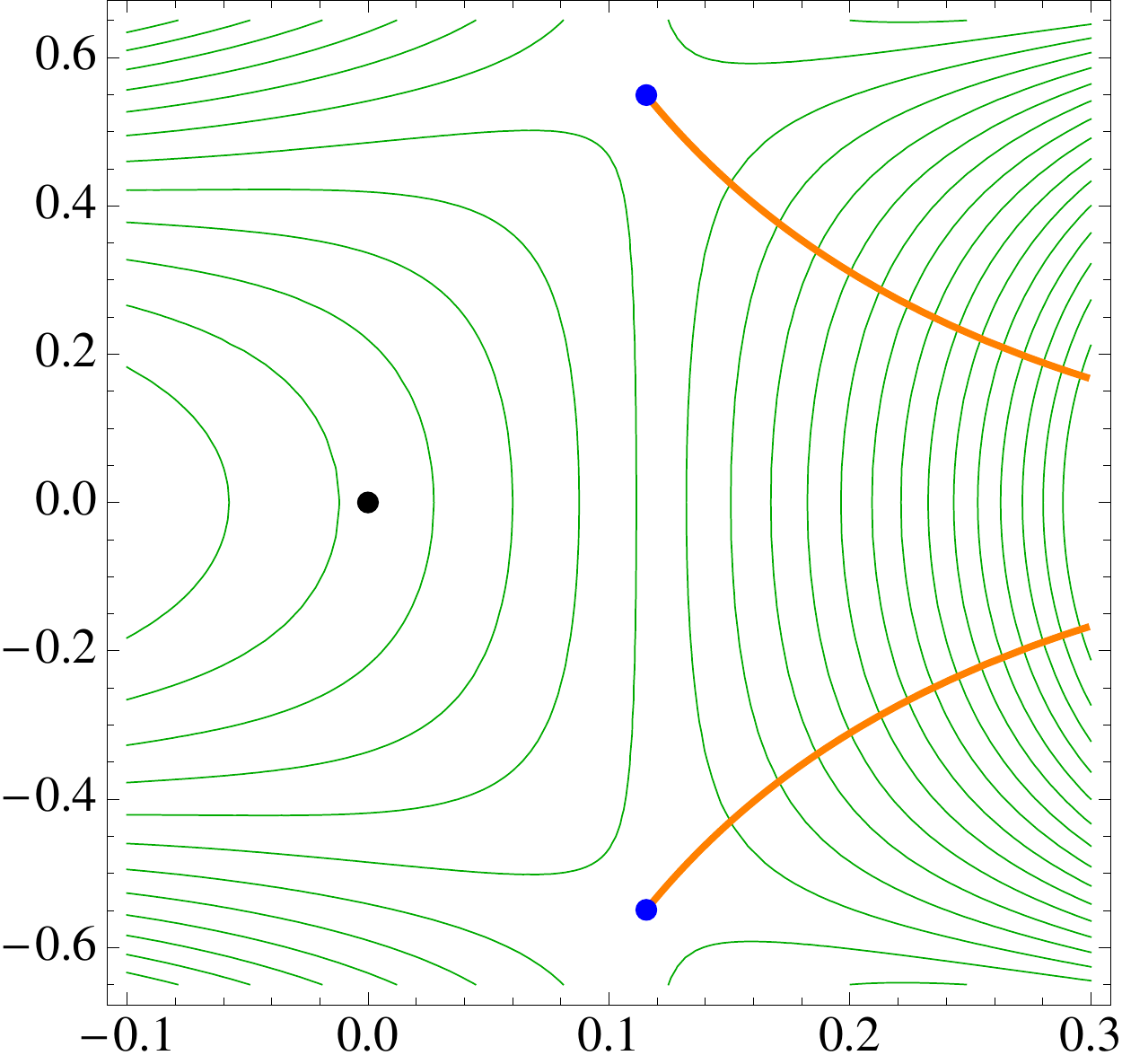}
\quad
\includegraphics[width=5cm]{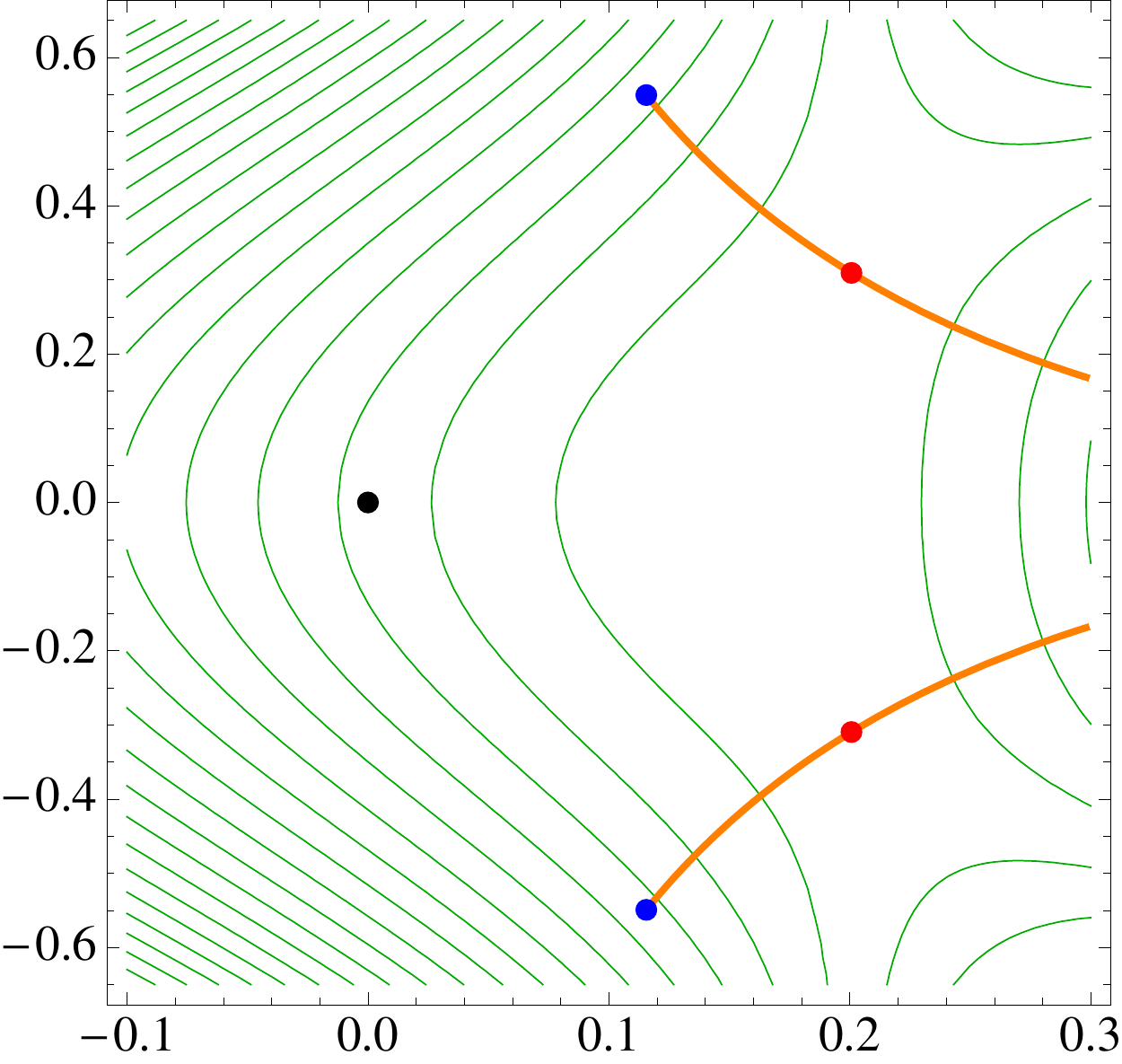}\quad
\includegraphics[width=5cm]{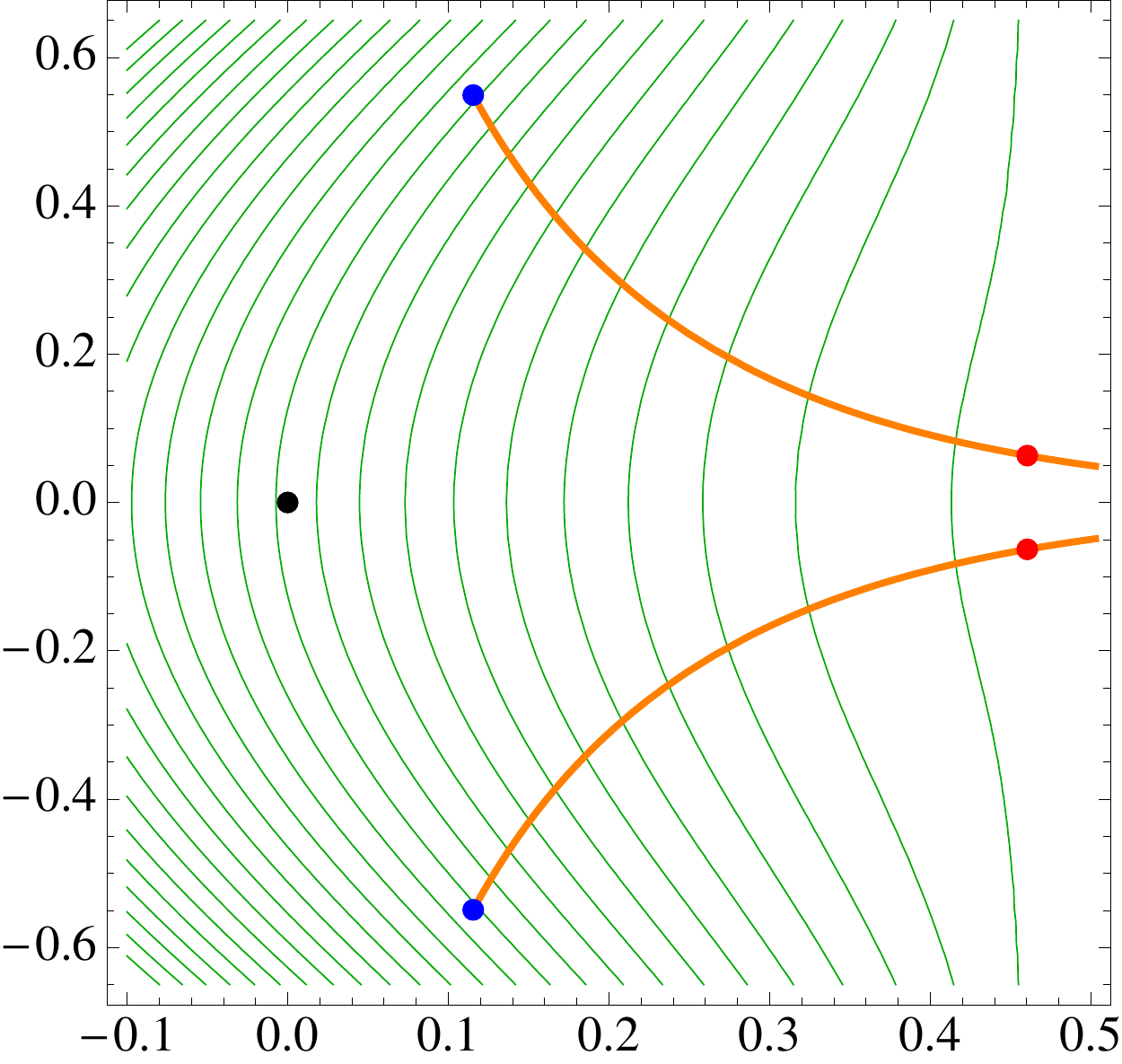}
\captionsetup{singlelinecheck=off}
\caption{
The superpotential $\cals V(\alpha,\beta,\chi,h)$ in the $(\alpha,\chi)$-plane with $\beta=\beta_{\frak a}$ and $h=h_{\frak a}$ kept constant at their critical values \eqref{AdS3sol} for  $ \frak a=0$,  $0.20$ and 0.248. The orange line   denotes the position of the $AdS_3\times \Sigma_{\frak g}$ critical points for $0\leq |\frak a|<1/4$. The end point denoted by the blue dot is the KPW point, the red dot is the critical point for the corresponding value of $\frak a$, and the black dot is the $SO(6)$ point. 
}
\label{SuperpotCont}
\end{figure}
%

It is straightforward to verify  that the critical points of $\mathcal{V}$, 
\begin{equation}\label{IRAdS3V}
\begin{gathered} 
\varphial  = \frac{1}{\sqrt{12}} \log\left(\frac{4}{1-16 \mathfrak{a}^2}\right) \;, \qquad \varphibe = \frac{1}{2} \log\left(\frac{1-4\mathfrak{a}}{1+4\mathfrak{a}}\right) \;,\\[6 pt]
\varphichi  = \sqrt{2} \log\left(\frac{2+\sqrt{1-16\mathfrak{a}^2}}{\sqrt{3+16\mathfrak{a}^2}}\right) \;,\qquad 
\varphih  = \frac{1}{\sqrt{6}} \log \left( \frac{1}{16 g^6} \frac{(3+16\mathfrak{a}^2)^3}{1-16 \mathfrak{a}^2}\right) \;,\end{gathered}
\end{equation}
are precisely the supersymmetric $AdS_3\times \Sigma_{\mathfrak{g}}$ vacua of interest \eqref{AdS3sol} shown in Figure~\ref{SuperpotCont}.  As is evident from the plots, the $SO(6)$ point (black dot) is not a critical point of $\cals V$ and the KPW point (blue dot) is a critical point only at $\frak{a}=0$, where the  blue and red dots coincide. The $AdS_3\times \Sigma_{\frak{g}}$ solutions in \eqref{AdS3a14} with $\frak{a}=\pm 1/4$ are not shown in Figure~\ref{SuperpotCont}.

 To construct the flow solutions, we will first examine  asymptotic expansions of \eqref{BPSeqnV} at both  UV points and in the IR.  In the special case of $\fa=0$, we will also find an analytic solution  for the flow  between the KPW point and  the corresponding $AdS_3\times \Sigma_\fg$ solution. For general $\fa$, solutions can be constructed only numerically and we will exhibit some of them.

\subsection{Asymptotic analysis}
\label{sec:asym}

The asymptotic analysis of the flow equations \eqref{rhodef} and \eqref{BPSeqnV} is quite similar to that for ordinary RG flows (see, e.g., \cite{Freedman:1999gp}) except that now there are two UV fixed points given by the two  $AdS_5$ solutions and a family of $AdS_3 \times \Sigma_{\frak{g}}$ fixed points in the IR labelled by $\mathfrak{a}$. For both UV fixed points, the corresponding asymptotically locally $AdS_5$ solutions   satisfy \cite{Maldacena:2000mw}
\begin{equation}\label{}
f~\sim ~h~\sim ~{\log{L\over r}}\,, \qquad  r~\longrightarrow~0\,,
\end{equation}
where $L$ is the $AdS_5$ radius given in \eqref{SO6fp} and \eqref{PWfp}. Using \eqref{rhodef}, we then have
\begin{equation}\label{asymfg}
f~\sim ~h~\sim {1\over 3}\,\rho\,,\qquad \rho~\sim~ {3\log{ L\over r} }~\longrightarrow~\infty\,.
\end{equation}
With the asymptotics of $f$ and $h$ fixed   by \eqref{asymfg}, it is convenient to rewrite the flow equations  \eqref{BPSeqnV} for the remaing fields using   
\begin{equation}\label{}
t~\equiv~ e^{-h}~\sim~ e^{-\rho/3}\,,
\end{equation}
as the independent variable. Setting
\begin{equation}\label{}
\varphi_i(t)\eql\varphi_i^{{ \rm UV}}+\phi_i(t)\,,\qquad i=\alpha,\beta,\chi\,,
\end{equation}
where $\varphi_i^{\rm UV}$ are the UV values of the scalar fields in \eqref{SO6fp} or 
\eqref{PWfp}, we obtain a  system of three first order equations of the form
\begin{equation}\label{fstorder}
t\,{d\phi_i\over dt}\eql A_i(t,\phi_\alpha,\phi_\beta,\phi_\chi)\,,\qquad i=\alpha,\beta,\chi\,.
\end{equation}
where $A_i(t,\phi_j)$ are holomorphic functions of $\phi_j$ and $t$ satisfying $A_i(0,0,0,0)=0$.

The asymptotic behavior for the solutions of interest can be obtained  as follows: First, we expand $A_i(t,\phi)$  to the linear order in the fields, $\phi_\alpha $, $\phi_\beta$ and $\phi_\chi $, and to the leading order in $t$. The resulting  linearized  system can be solved analytically and  its solution   determines   the structure of the local series expansion for  the exact solutions to the nonlinear system, see for example \cite{Hukuhara}. 

At the $SO(6)$ point  \eqref{SO6fp},  the linearization of \eqref{fstorder} gives
\begin{equation}\label{theeqs1}
t\,{d\phi_\alpha \over dt}\eql {t^2\over\sqrt 3\,g^2}+{2}\,\phi_\alpha \,,
\qquad
t\,{d\phi_\beta\over dt}\eql {4\frak a\over g^2}\,t^2+ {2}\,\phi_\beta\,,\qquad 
t\,{d\phi_\chi \over dt}\eql\phi_\chi  \,.
\end{equation}
The general solution to \eqref{fstorder} at the $SO(6)$ point can then be obtained by expanding the three fields into power series in $t$ and $t^2\log t$.  The resulting recurrence for the expansion coefficients  is consistent and yields the following general solution:  
\begin{equation}\label{sersolso6}
\begin{split}
 \phi_{\alpha} (t) & \eql \Big({1\over \sqrt 3\,g^2}-{2\over\sqrt 3}\,c_0^2\Big)\,t^2\log t+a_0\,t^2
 +\ldots \,,\\[6 pt]
 \phi_{\beta}(t)& \eql {4\frak a\over g^2}\,t^2\log t+b_0\,t^2+\ldots\,,\\[6 pt]
 \phi_{\chi} (t) & \eql c_0\,t+\Big({4c_0^3\over 3}-{2c_0\over 3 g^2}\Big)\,t^3\log t+\Big({c_0\over g^2}-{2\,a_0c_0\over\sqrt 3}-{7\,c_0^3\over 12}\Big)\,t^3+\ldots\,,
\end{split}
\end{equation}
where the coefficient of the omitted higher order terms are completely determined by  $a_0$, $b_0$, $c_0$, and  $\fa$. Note that the leading terms in \eqref{sersolso6} can be obtained as an exact solution to the linearized system \eqref{theeqs1}.

We note that all solutions \eqref{sersolso6} vanish as $t\rightarrow 0$. This means that the $SO(6)$ point should act as a local ``attractor point'' in the UV, in the sense that a generic flow solution  will asymptote to that point  as $\rho\rightarrow\infty$. We will see that this  expectation is indeed confirmed by the numerical results below.

The  other $AdS_5$ solution is the KPW point \eqref{PWfp}. The linearization of \eqref{fstorder} around this point gives the following  equations
\begin{equation}\label{linpweqs}
\begin{split}
t\,\dfrac{d\phi_{\alpha}}{dt} &=  2\,\phi_{\alpha}   -\sqrt{6}\,\phi_{\chi}\,,\qquad 
t\,\dfrac{d\phi_{\beta}}{dt}  = {3\cdot 2^{2/3} \,\frak a\over g^2}\,t^2+2\,\phi_{\beta}\;,\qquad 
t\,\dfrac{d\phi_{\chi}}{dt} = -{\sqrt 6}\,\phi_{\alpha} \;,
\end{split}
\end{equation}
which are solved by 
\begin{equation}\label{LSsersol}
\begin{split}
\phi_\alpha (t) & \eql p_0\,t^{1+\sqrt 7}+s_0\,t^{1-\sqrt 7}\,,\\[6 pt]
\phi_\beta(t) & \eql {3\cdot 2^{2/3}\over g^2}\,{\frak a }\,t^2\log t+q_0\,t^2\,,\\[6 pt]
\phi_\chi (t) & \eql {1-\sqrt 7\over\sqrt 6}\,p_0\,t^{1+\sqrt 7}+{1+\sqrt 7\over\sqrt 6}\,s_0\,t^{1-\sqrt 7}
\,.
\end{split}
\end{equation}
The general solution to \eqref{fstorder} that vanishes as $t\rightarrow 0$ can be found as a power series in  $t^2$, $t^2\log t$  and  $t^{1+\sqrt 7}$ with the leading terms given in \eqref{LSsersol} with $s_0=0$. The  subleading terms, which we omit here,  have coefficients fixed in terms of $p_0$, $q_0$, and $\mathfrak{a}$.

Using the standard holographic dictionary, the expansions \eqref{sersolso6} and \eqref{LSsersol} are in perfect agreement with the  field theory picture in Section~\ref{Sec:FieldTheory}. The operators $\cals O_\alpha$, $\cals O_\beta$ and $\cals O_\chi$ in  $\mathcal{N}=4$ SYM, see \eqref{N=4ops} and \eqref{Ochi},   are dual to the supergravity fields, $\alpha$, $\beta$ and $\chi$, and at the $SO(6)$ point have dimensions 2, 2, and 3, respectively. These  are consistent   with \eqref{sersolso6}, where the most singular terms determine the sources and the subleading terms determine the expectations values for the corresponding  operators.

As expected,  we read off from \eqref{sersolso6} that a nontrivial  background flux  amounts to turning on  sources for the  bosonic bilinears, $\cals O_\alpha $ and $\cals O_\beta $, in  $\cN=4$ SYM. In particular,  the source for $\cals O_\beta $ depends on the magnitude of the background flux, $\fa$, while that for   $\cals O_\alpha$  is constant. The latter  can be traced to the constant coefficient of the background flux along  the $R$-symmetry generator in \eqref{sugraR} and that in turn follows from the  particular solution of \eqref{acond} in \eqref{aconstr} we have chosen. As for the untwisted RG flows \cite{Freedman:1999gp}, the parameter $c_0$ in \eqref{sersolso6} is proportional to the source for the operator $\cals O_\chi$, while the parameters $a_0$ and $b_0$ are related to the vevs for the bosonic bilinear operators  $\cals O_\alpha$ and $\cals O_\beta$ in \eqref{N=4ops}.

At the KPW point,  from \eqref{LSsersol} we have three operators, $\cals O_\Delta$, of dimension   $\Delta=2$, $1+\sqrt 7$ and $ 3+\sqrt 7$, respectively. The operator $\cals O_2$ 
is dual to the scalar $\beta$, while a relevant operator $\cals O_{1+\sqrt 7}$ and an irrelevant operator $\cals O_{3+\sqrt 7}$ are dual to linear combinations of the scalars  $\alpha$ and $\chi$.
We see that as before the background flux sources the operator, $\cals O_2$,  with the overall coefficient determined by the cosmological constant of the $AdS_5$ solution. The absence of a constant source, which was present at the $SO(6)$ point, is consistent with the uniqueness of the $R$-symmetry current at the LS fixed point. 

It may seem surprising at first that in the linearized expansion  \eqref{LSsersol} there is no term of the form $t^{3-\sqrt{7}}$, which would correspond to a source of the relevant operator  $\cals O_{1+\sqrt 7}$ in the dual field theory. The reason for  the absence of such a source is that the operators $\cals O_{1+\sqrt 7}$ and $\cals O_{3+\sqrt 7}$ lie in the same (unprotected) massive vector supermultiplet in the LS fixed point  (see, Table 6.2 in \cite{Freedman:1999gp}). Therefore, if we turn off the source for the operator $\cals O_{3+\sqrt 7}$ by setting $s_0=0$, then we must also  turn off the source for the operator $\cals O_{1+\sqrt 7}$ to ensure that   supersymmetry is preserved.\footnote{One can use the same argument to explain the observation in  \cite{Khavaev:2000gb,Khavaev:2001yg} that holographic RG flows out of the KPW point with $\mathbb{R}^{1,3}$ slicing involve only vevs for the operator $\cals O_{1+\sqrt 7}$. } This amounts to being able to turn on only a vev for the operator $\cals O_{1+\sqrt 7}$, which is proportional to  $p_0$ in \eqref{LSsersol}. As usual, the parameter $q_0$ is related to the vev for the operator $\mathcal{O}_2$. In summary, the supersymmetric RG flow away from the LS fixed point is driven by the operator $\cals O_2$ sourced by the background flavor flux proportional to $\mathfrak{a}$. 

%
\begin{figure}[t]
\centering
\includegraphics[width=7cm]{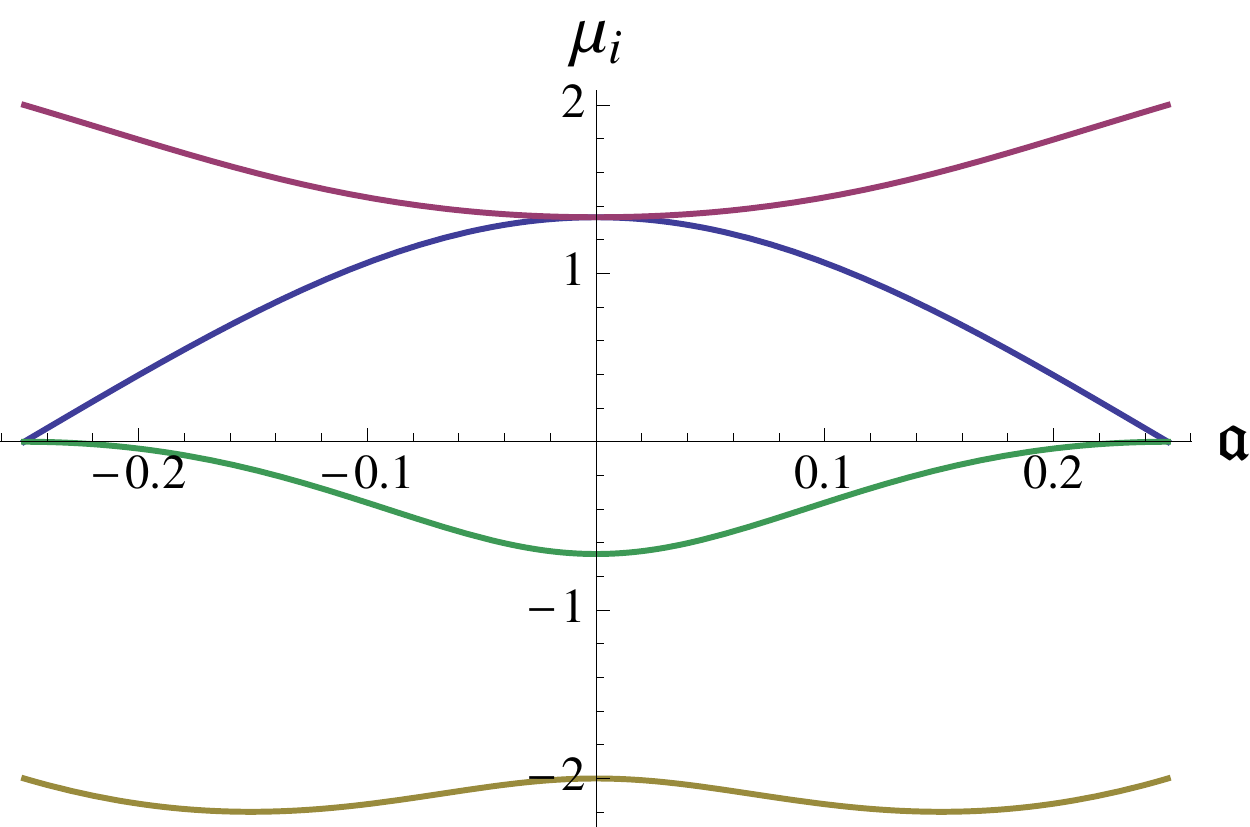}
\caption{Eigenvalues of the mass matrix $M_{ij}$.
}
\label{figeigs}
\end{figure}
%

Let us now turn to   the IR region close to the $AdS_3\times \Sigma_{\frak g}$ solutions. At the critical values  $\varphi_i=\varphi^{\rm IR}_i$ given in   \eqref{IRAdS3V}, we find 
\begin{equation}\label{}
\cals V_{\rm IR}\equiv \cals V(\varphi^{\rm IR})\eql -{2\,g^3\over 3+16\,\fa^2}\,,
\end{equation}
and thus 
\begin{equation}\label{}
\rho \sim -\log(|\cals V_{\rm IR}|\,r)\quad \longrightarrow\quad -\infty\,.
\end{equation}
Setting $
\varphi_i = \varphi_i^{\rm IR}  + \phi_i$ and expanding \eqref{BPSeqnV} to the leading order, we obtain a linear system
\begin{equation}\label{linMeqs}
{d\phi_i\over d\rho} = M_{ij} \phi_j\,,\qquad i,j=\alpha,\beta,\chi,h\,,
\end{equation}
where $M_{ij}=M_{ji}$ is a symmetric ``mass matrix'' with the following nonvanishing entries 
\begin{equation}\label{IRmassmatrix}
\begin{gathered}
M_{11} \eql -{2\,(3+80\fa^2)\over 3\,(3+16\,\fa^2)}\,,\qquad M_{12}\eql -{32\fa\over \sqrt 3\,(3+16\,\fa^2)}\,,\qquad M_{13}\eql {2\sqrt 6\sqrt{1-16\,\fa^2}\over 3+16\,\fa^2}\,, \\
M_{14}  \eql -{64\sqrt 2\,\fa^2\over 3(3+16\,\fa^2)}\,,\qquad M_{22}\eql -{2( 1-16\,\fa^2)\over  3+16\,\fa^2}\,,\qquad M_{23}\eql {8\sqrt 2\,\fa\,\sqrt{1-16\,\fa^2}\over 3+16\,\fa^2}\,,\\
M_{24} \eql {16\sqrt 2\,\fa\over\sqrt3\,(3+16\,\fa^2)}\,,\qquad M_{44}\eql {4\over 3}\,. 
\end{gathered}
\end{equation}

Note that the  mass matrix depends on the background flux, $\fa$, but does not depend on~$g$. 
For $\fa=0$, its eigenvalues, $\mu_i$, are $4/3, 4/3, -2/3$, and $-2$,  and the same  pattern of two positive and two negative eigenvalues persists throughout the whole  range $|\fa| < 1/4$ as shown in Figure~\ref{figeigs}. This means that for a fixed value of $\mathfrak{a}$ in the IR, we should have a two-parameter family of flows into the $AdS_3$ point tangent to the plane spanned by the eigenvectors, $v_{(1)}$ and $v_{(2)}$, of the mass matrix for the two positive eigenvalues. For the special values $\frak{a} = \pm 1/4$, two of the eigenvalues of $M_{ij}$ vanish. This is consistent with the explicit solution   \eqref{AdS3a14}, where the scalar $\chi$ vanishes and the scalar $\alpha$ is a modulus. The corresponding fluctuations are the two zero modes.

\subsection{Analytic example}

\begin{figure}[t]
\centering
\includegraphics[height=4.35cm]{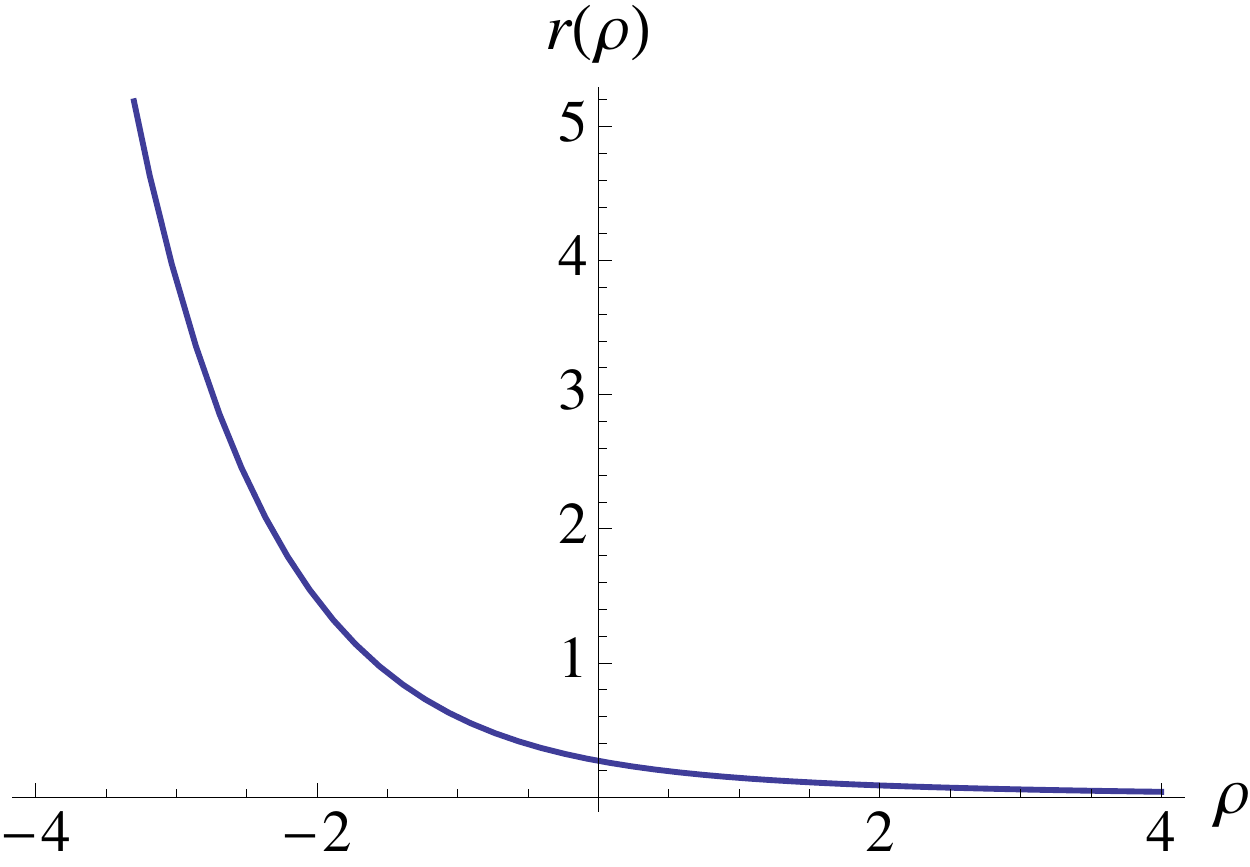}
\qquad
\includegraphics[height=4.5cm]{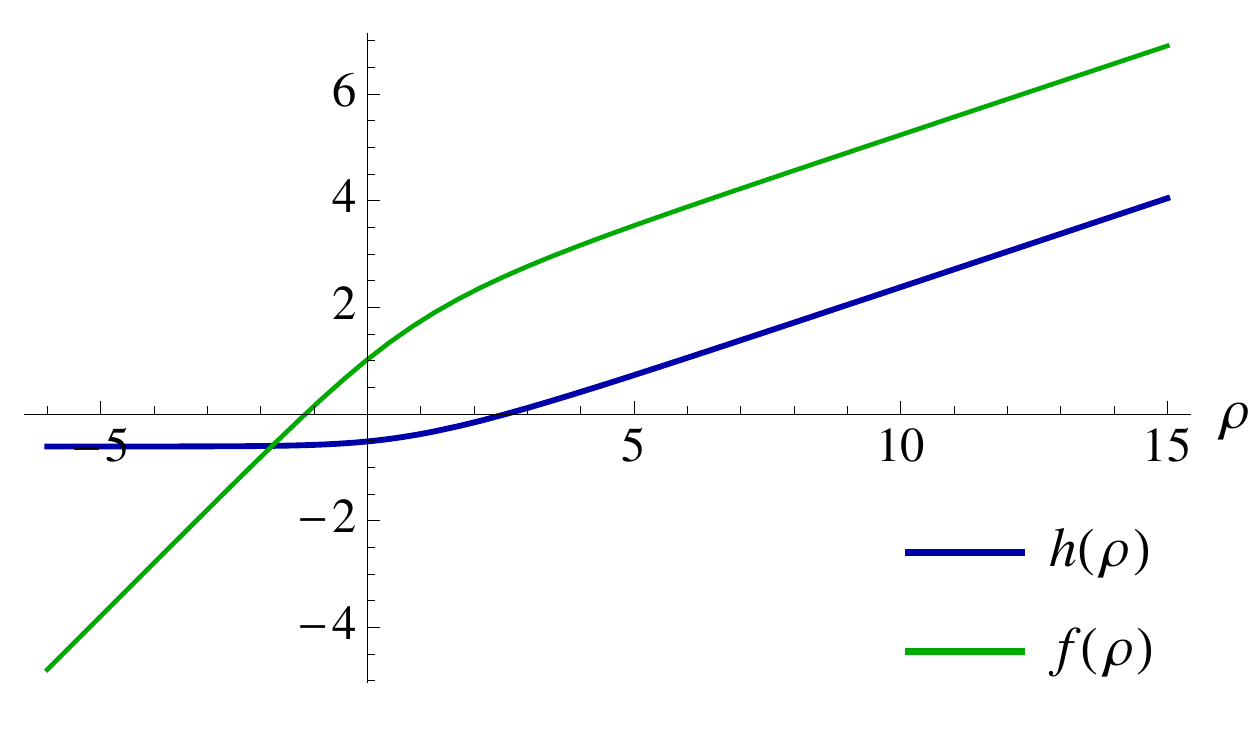}\\[10 pt]
\includegraphics[height=5.75cm]{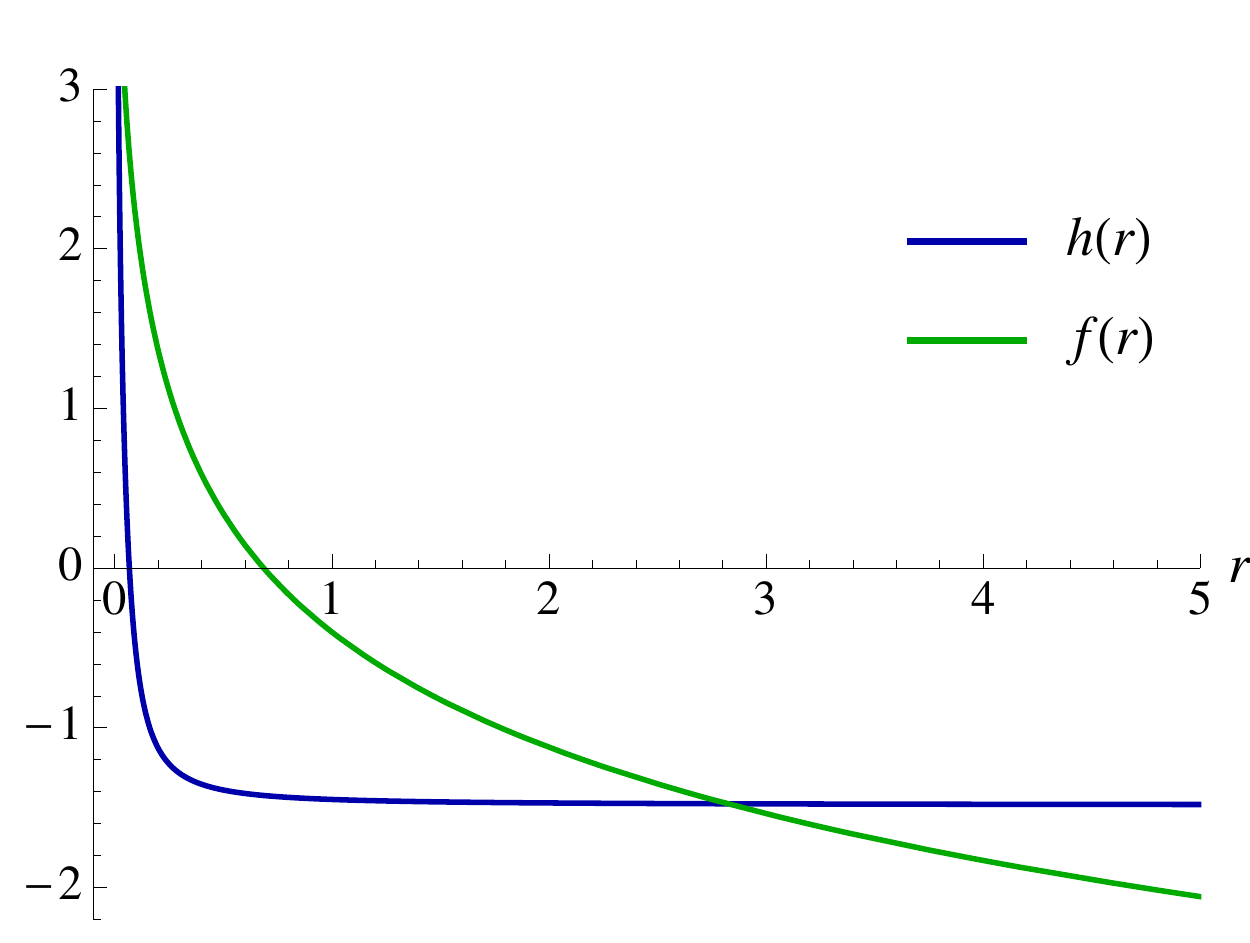}
\caption{The radial variable and the two metric functions $f$ and $h$ for the analytic solution in \eqref{analytic} and \eqref{rrho}.
}
\label{figanalytic}
\end{figure}
%

In general it is not possible to solve the system of equations \eqref{BPSeqnV} analytically.  There is, however, a special value of the parameter, $\frak a$,  namely  $\mathfrak{a}=0$, for which the flow equations admit the following simple analytic solution,
\begin{equation}\label{analytic}
\begin{split}
\varphial  &= \frac{\log 2}{\sqrt{3}}\;, \qquad \varphibe = 0\;, \qquad \varphichi  = \frac{\log 3}{\sqrt{2}}\;, \\[6 pt]
\varphih  &= \frac{1}{\sqrt{6}}\log\Big[\,\frac{3^3}{2^7g^6} \Big(\sqrt{c^2 e^{4\rho/3}+1}+1\Big)^3\,\Big]\;.
\end{split}
\end{equation}
Here $c$ is an integration constant, which can be set to one by a constant shift of the radial variable $\rho \to \rho - \frac{3}{2}\log c$.  
Substituting the solution \eqref{analytic} in \eqref{rhodef}, we find the following explicit relation between the two radial coordinates
\begin{equation}\label{rrho}
r(\rho)\eql {3\over 8\,g^3}\Big[2 e^{-\rho } \big(\sqrt{e^{4 \rho /3}+1}+1\big) -e^{\rho /3} \, _2F_1\left(\coeff{1}{4},\coeff{1}{2};\coeff{5}{4};-e^{4 \rho /3}\right) +\coeff{1}{\sqrt \pi}\Gamma (\coeff{1}{4}) \Gamma (\coeff{5}{4})\Big]\;.
\end{equation}
In Figure \ref{figanalytic}, we have plotted $r(\rho)$, together with    $h$ and $f$ as functions of both $r$ and $\rho$. Note that the asymptotic behavior of those functions agrees with \eqref{asymfg}, \eqref{IRAdS3V} and \eqref{AdS3sol}.

The solution \eqref{analytic} is special in that the supergravity scalars remain fixed at  their values for the $AdS_5$ solution at the KPW point \eqref{PWfp}. Thus the only quantities that change along the flow are the metric functions $f(r)$ and $h(r)$. There is a similar analytic supersymmetric flow solution which interpolates between the $SO(6)$ point and an $AdS_3 \times \Sigma_{\frak{g}}$ vacuum \cite{Maldacena:2000mw}. 
In fact, as discussed in \cite{Benini:2013cda,BBC}, any minimal five-dimensional gauged supergravity admits such an analytic flow solution. This solution should describe a universal RG flow, triggered by a twisted compactification on a Riemann surface, in any $\mathcal{N}=1$ SCFT with a holographic dual.

\subsection{Numerical solutions}

\begin{figure}[t]
\centering
\includegraphics[width=7cm]{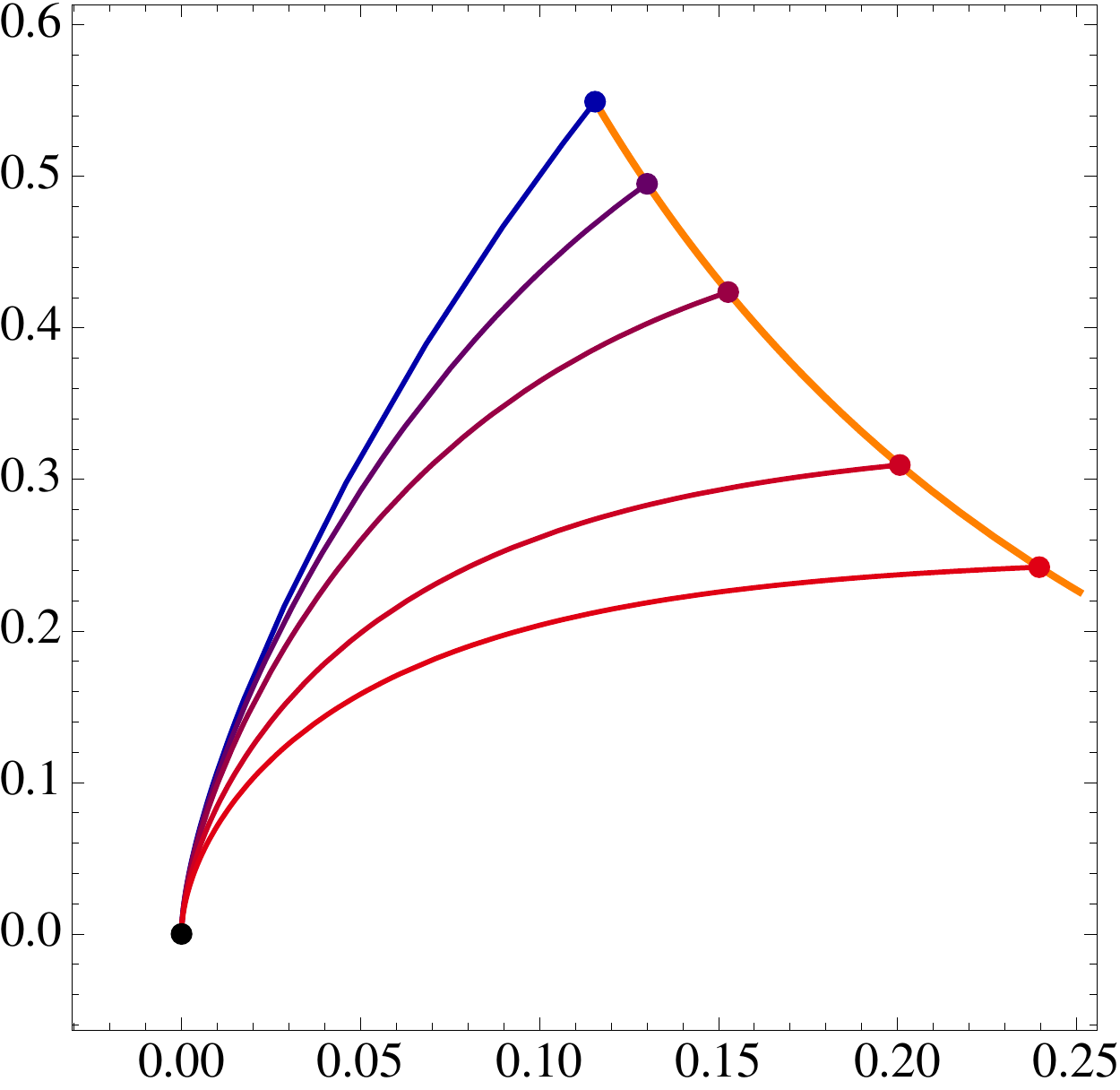}
\caption{Examples of flows  from the $SO(6)$ point to different $AdS_3\times \Sigma_{\frak g}$ solutions projected onto the $(\alpha,\chi)$-plane for  $\frak a=0.001,\,0.1,\, 0.15,\, 0.2 $, and $0.22$ from left to right. 
}
\label{SO6Traj}
\end{figure}
%
For an arbitrary value of the background flux, $\fa$, the flow equations \eqref{BPSeqnV} can only  be solved numerically. In the following, we construct some representative    solutions for different classes of RG flows predicted by the field theory analysis in Section~\ref{Sec:FieldTheory}.  

As usual, we find that the   integration of the first order system \eqref{BPSeqnV} is numerically more stable if we specify the initial conditions in the IR close to an $AdS_3\times\Sigma_{\frak g}$ critical point. Hence, for a given $|\frak a| < 1/4$, we take $\rho_0\ll 0$ and set
\begin{equation}\label{initcond}
\varphi_i(\rho_0)=\varphi_i^{\rm IR}+  \phi_i^{(0)}\,, \qquad \phi_i^{(0)}\eql \xi^{(1)}\,v_{(1)}^i+\xi^{(2)}\,v_{(2)}^i\,, \qquad
 i=\alpha,\,\beta,\,\chi,\,h\,,
\end{equation}
where  $v_{(1)}$ and $v_{(2)}$ are  two orthonormal  eigenvectors for the positive eigenvalues of the  mass matrix  in  \eqref{linMeqs} and 
$\xi^{(1)}$ and $\xi^{(2)}$ are arbitrary  small parameters. In the examples below, we typically work with $\rho_0\sim -10$ and $|\xi^{(1,2)}|\sim 10^{-4}$. Since the other two eigenvalues of the mass matrix are negative, this choice of initial conditions does not guarantee numerical stability as we integrate \eqref{BPSeqnV} towards the $AdS_3$ critical point where $\rho\ll \rho_0$. However, by extrapolating the linearized analysis to  full nonlinear solutions, it is reasonable to assume  that any flow from an $AdS_3$ critical point is asymptotic for $\rho\gg \rho_0$ to a solution in the class   we are considering.

\begin{figure}[t]
\centering
\includegraphics[width=7cm]{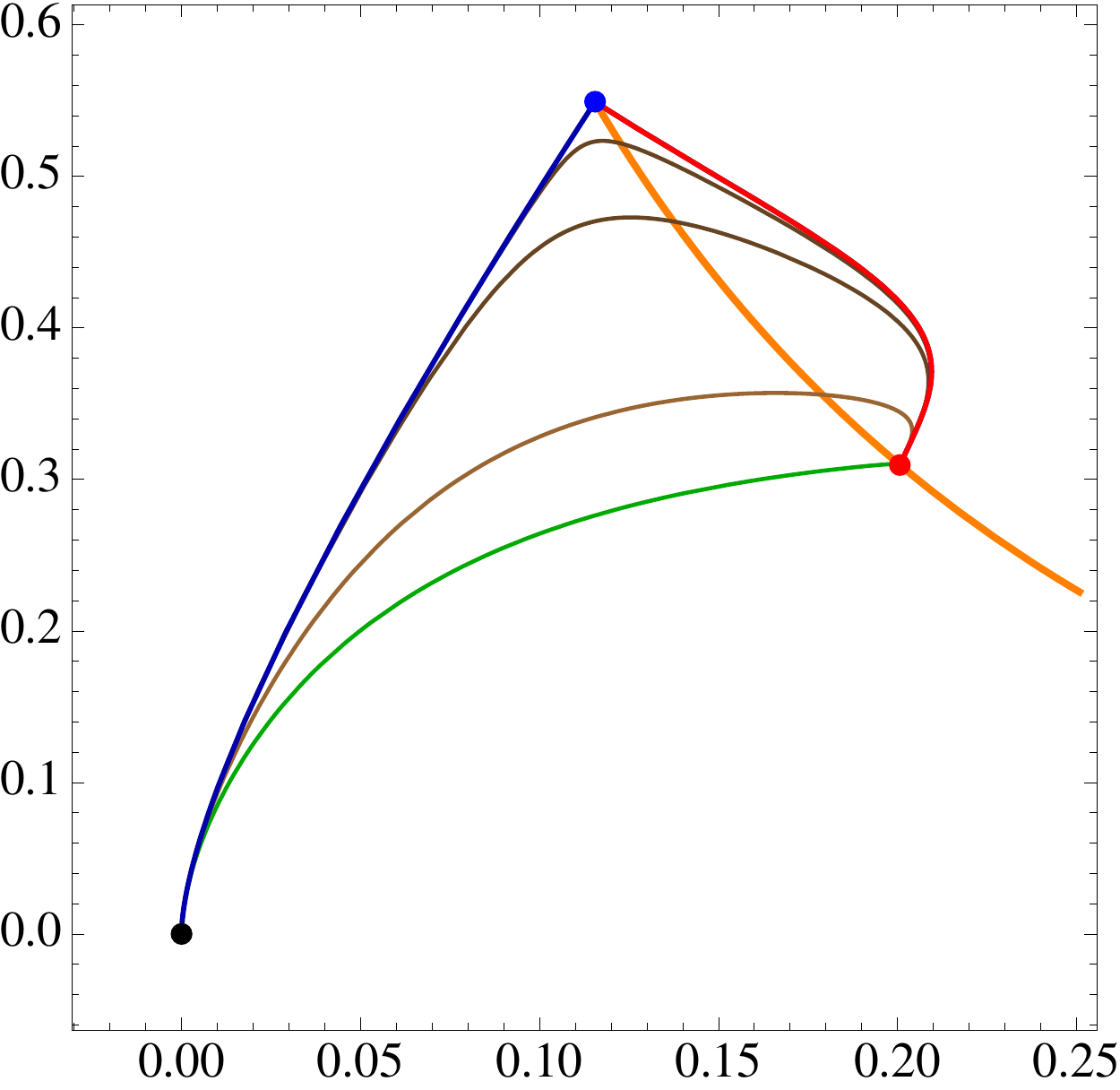}
\caption{Solutions of the flow equations \eqref{BPSeqnV} projected onto the $(\alpha,\chi)$-plane for $\frak a=0.20$. The   curves between the $SO(6)$ point (black dot) and the $AdS_3\times \Sigma_{\frak{g}}$ point (red dot) are representatives of the one-parameter family of holographic RG flows labelled by the parameter $c_0$ in \eqref{sersolso6} corresponding to the mass $m$ in \eqref{LSsuperpot}. The red curve connecting the blue and red dots is the unique holographic RG flow between the KPW point and the $AdS_3\times \Sigma_{\frak{g}}$ solution.
}
\label{SuperpotTraj}
\end{figure}

As one might have expected from the asymptotic analysis in Section~\ref{sec:asym}, a generic solution for small $\xi^{(1,2)}$, or equivalently, small   velocites in the IR, remains in the basin of attraction of the $SO(6)$ critical point in the UV.  In particular,   for a given $AdS_3\times \Sigma_{\frak g}$ critical point we find a one-parameter family of flows  that originate at  the $SO(6)$ point. Examples of such flows projected onto the $(\alpha,\chi)$-plane are shown in Figures~\ref{SO6Traj} and \ref{SuperpotTraj}. This family can be parametrized by $c_0$ in \eqref{sersolso6}, which in turn corresponds to the mass $m$ in the LS superpotential \eqref{LSsuperpot}. This agrees with the field theory expectation   discussed in Section~\ref{subsec:N=4flowsFT}.

%

The flows from the KPW point in the UV are more subtle. First, we are looking for   solutions  in the four-dimensional space of fields, $\alpha$, $\beta$, $\chi$ and  $h$, that  interpolate  between two points, an $AdS_3$ critical point and the KPW point, where the latter has    $h\rightarrow\infty$.   Since we have only two tunable parameters, $\xi^{(1)}$ and $\xi^{(2)}$,  to work with, the  existence of such solutions is by no means guaranteed. Secondly, unlike the $SO(6)$ point, the KPW point is numerically unstable, because of the presence of the    $t^{1-\sqrt 7}$ mode  in the linearized solution \eqref{LSsersol}. Therefore, any numerical flow trajectory obtained by shooting from the IR will eventually start  moving away from the KPW point.  Hence, all we can hope for is to see some numerical evidence that,  by fine tuning of the initial conditions, one can obtain solutions  which remain close to the KPW point within a large range of $\rho\gg 0$. Indeed, this is precisely what we find.

%
\begin{figure}[t]
\centering
\includegraphics[width=7cm]{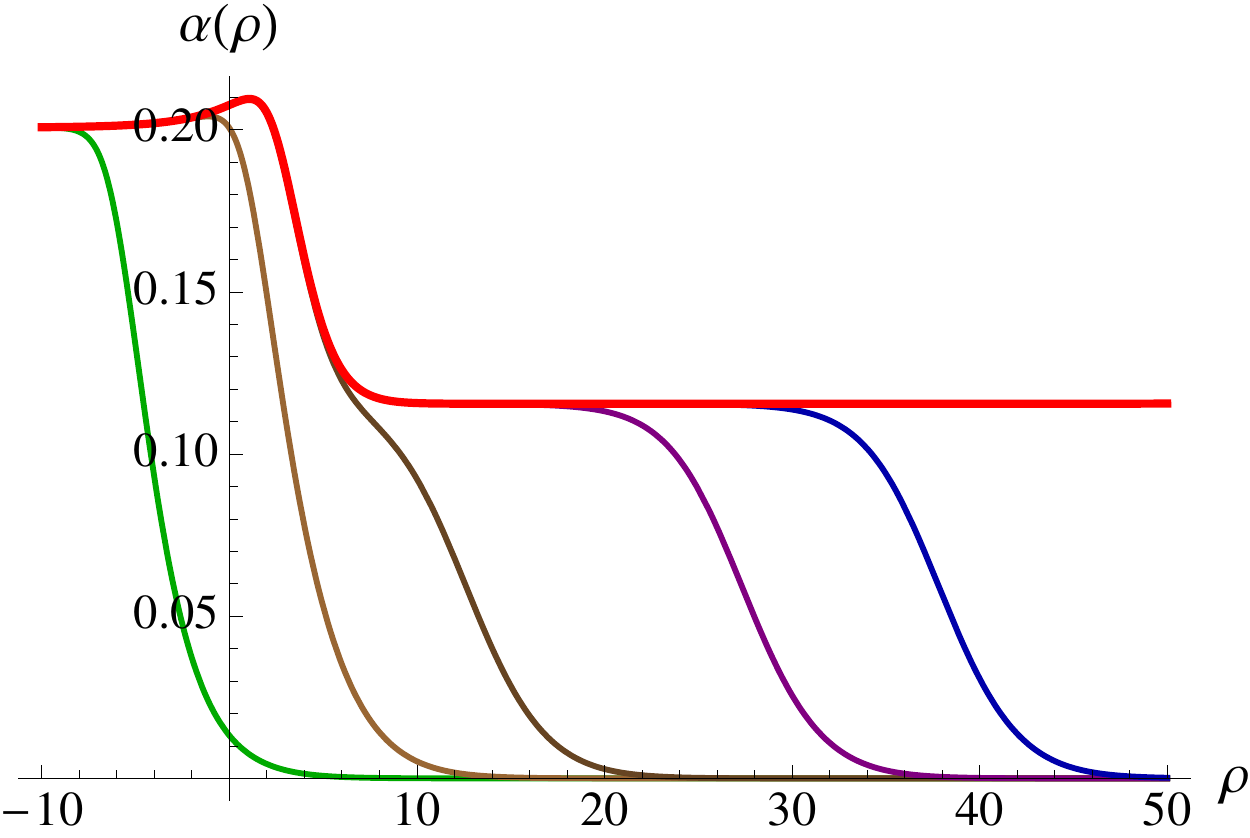}
\qquad
\includegraphics[width=7cm]{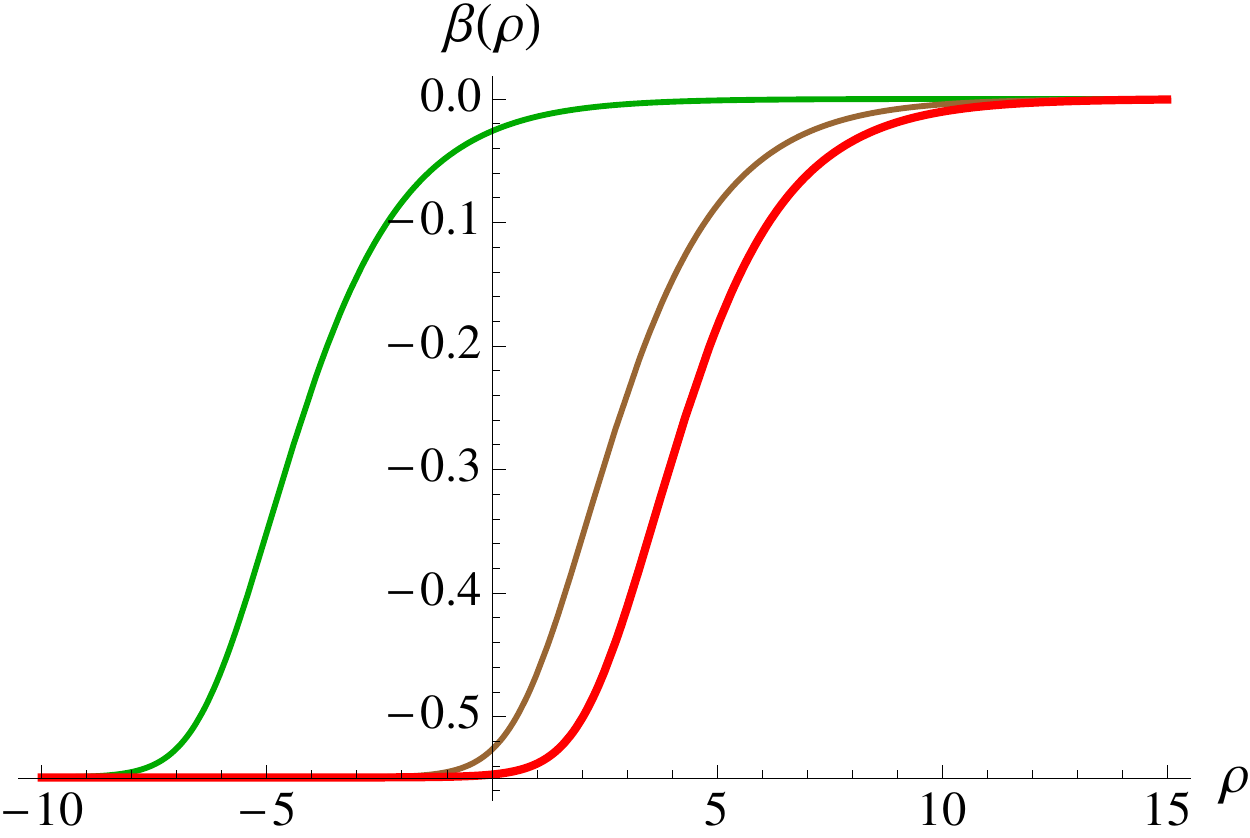}\\[12 pt]
\includegraphics[width=7cm]{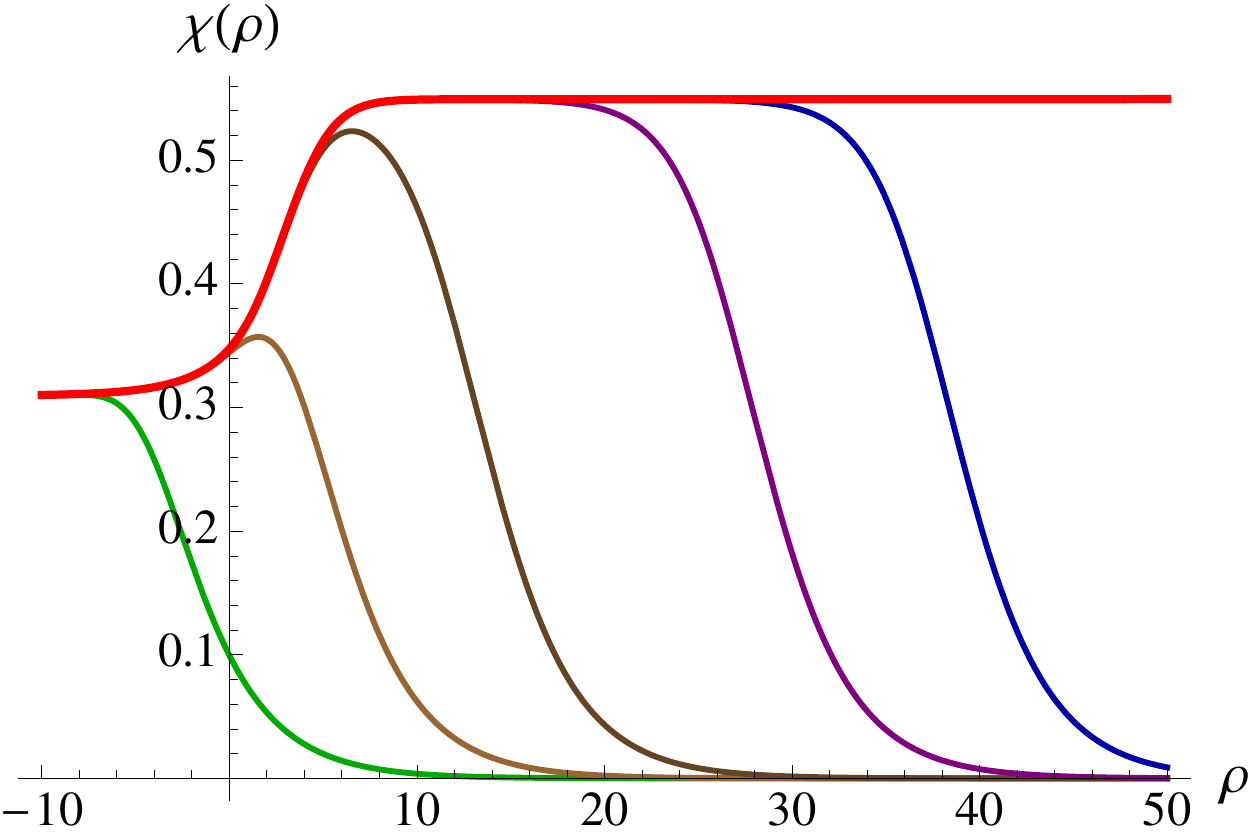}
\qquad
\includegraphics[width=7cm]{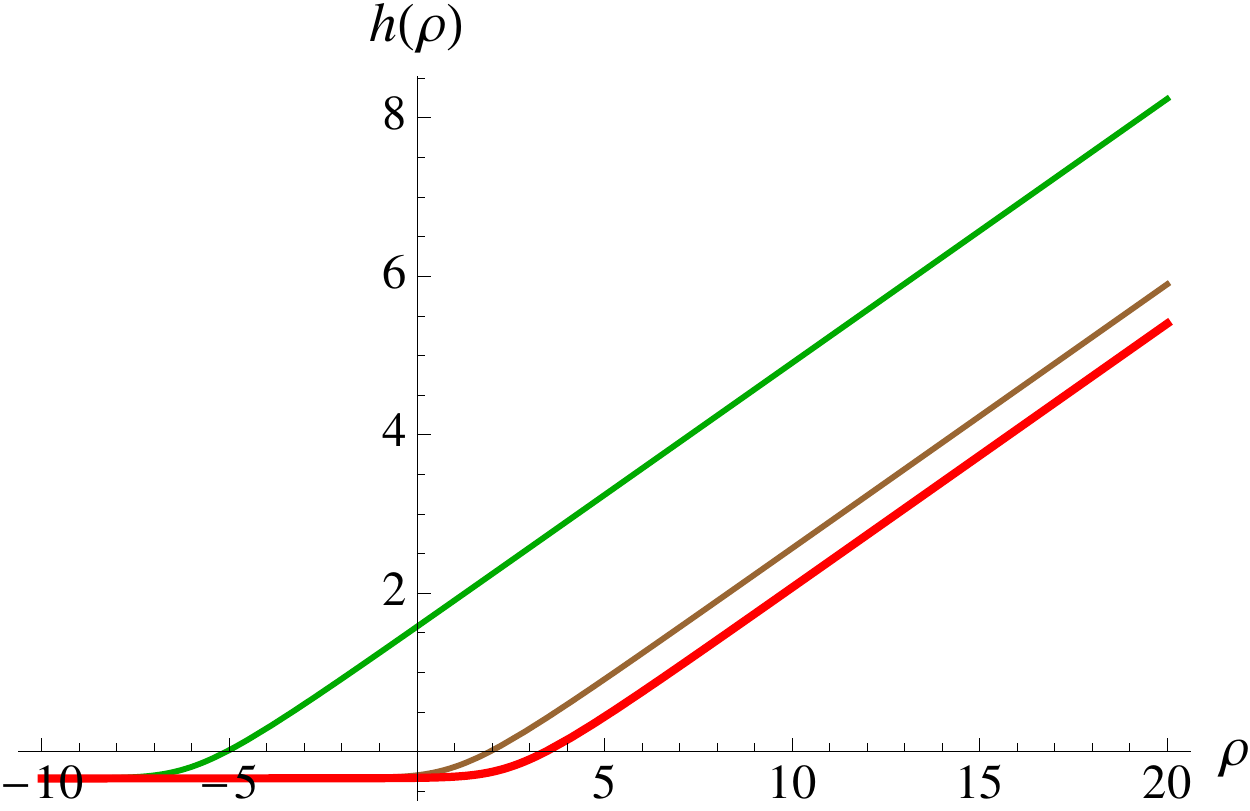}\\
\caption{
A family of solutions of the flow equations \eqref{BPSeqnV} for $\mathfrak{a}=0.20$ illustrating the tuning of initial conditions for the RG flow (red plot) from the KPW point in the UV to an $AdS_3\times \Sigma_{\frak g}$ point in the IR.  The colors of the plots are correlated with those  in  Figure \ref{SuperpotTraj}.}
\label{FamSol}
\end{figure}

A typical family of flow solutions for $\frak{a}=0.20$ is presented in Figure~\ref{SuperpotTraj}.  One should keep in mind that the plot represents a projection from the four-dimensional space of scalar fields onto the $(\alpha,\chi)$-plane. The family of solutions is constructed by varying the parameters $\xi^{(1)}$ and $\xi^{(2)}$ in \eqref{initcond}. As is clear from Figure~\ref{SuperpotTraj}, generically the flow is to the $SO(6)$ point. However, by fine tuning of the parameters $\xi^{(1)}$ and $\xi^{(2)}$, the UV side of the projection of the flow onto the $(\alpha,\chi)$-plane  can be brought arbitrarily close  the KPW point as illustrated in Figure~\ref{SuperpotTraj}.  The existence of such a family of flows  might  not seem that  surprising given that we have two free parameters and the tuning is done in a plane of two scalar fields. It turns out, however, that by bringing the projection of the flow close to the KPW point in the $(\alpha,\chi)$-plane,  the flow itself in the four-dimensional space of scalar fields approaches the $AdS_5$ solution for the KPW point. This is a nontrivial evidence that this finely tuned flow indeed exists. The behavior of all four scalars in the family of flow solutions with $\frak{a}=0.20$ is shown in Figure~\ref{FamSol}, whose colors are correlated with those in Figure~\ref{SuperpotTraj}. Note that the curves in Figure~\ref{FamSol} that appear missing in the plots for $\beta(\rho)$ and $h(\rho)$ lie too close to the ones that are shown to be visible.

The red curves represent  our best approximation to the holographic RG flow from the KPW point to the $AdS_3\times \Sigma_{\frak{g}}$ point at $\frak{a}=0.20$ that we are looking for. Similar flows can be constructed also for other values of the parameter $\frak{a}$ in the range $|\frak{a}|<1/4$.\footnote{Holographic RG flows between the $SO(6)$ point and the $(2,2)$ $AdS_3\times \Sigma_{\frak{g}}$ vacua for $\frak{a}=\pm 1/4$ were constructed numerically in \cite{Benini:2013cda}. There are no flows from the KPW point to the $AdS_3\times \Sigma_{\frak{g}}$ solutions for $\frak{a}=\pm 1/4$.} The flow solutions we have exhibited should be considered as a strong evidence from holography that the SCFTs described in Section \ref{Sec:FieldTheory} indeed exist and the RG flows to them are dynamically realized. 

\sect{Conclusions}
\label{Sec:Conclusions}

In this paper we have provided ample evidence, both in field theory and in supergravity,  that there exists a family of two-dimensional SCFTs  obtained by a twisted compactification of the $\mathcal{N}=1$ LS SCFT on a closed Riemann surface $\Sigma_{\frak{g}}$. The existence of this family of SCFTs and their dual $AdS_3\times \Sigma_{\frak{g}}$ solutions and holographic RG flows raises a number of interesting points.

It is known that non-Abelian flavor symmetries cannot mix with the $R$-symmetry  in purely four-dimensional RG flows \cite{Intriligator:2003jj}. However, when one places the four-dimensional theory on a Riemann surface and turns on a background gauge field  for the Cartan subgroup of the non-Abelian flavor symmetries, the infrared dynamics is richer. The  flavor flux provides a parameter which labels different two-dimensional IR fixed points. The superconformal $R$-symmetry undergoes nontrivial mixing during an RG flow from four dimensions and, at the two-dimensional fixed point, it becomes  a linear combination of the original $R$-symmetry and a subgroup of the flavor symmetry. The same scenario is present in the RG flows from six to four dimensions discussed in \cite{Bah:2011vv,Bah:2012dg}. Thus we can conclude  that turning on background magnetic fluxes for flavor symmetries  provides a general mechanism for constructing families of interacting SCFTs.

It would be desirable to have  a two-dimensional field theory construction for the family of interacting $(0,2)$ SCFTs found in this paper.   One possible route is to study twisted compactifications of $\mathcal{N}=4$ SYM or the $\mathcal{N}=1$ LS SCFT on punctured Riemann surfaces. In this way one may be able to  identify a SCFT  that would serve as an elementary ``building block" for  more general SCFTs corresponding to closed Riemann surfaces. Such a construction would be analogous to the one for four-dimensional SCFTs of class $\mathcal{S}$ \cite{Gaiotto:2009we,Bah:2011vv,Bah:2012dg}. It would also provide  insights into the rich set of dualities (or trialities \cite{Gadde:2013lxa}) that are expected to exist for these two-dimensional theories. 

On the supergravity side, it would be interesting to uplift  our $AdS_3\times \Sigma_{\frak{g}}$ vacua to solutions of type IIB supergravity in ten dimensions. Most likely, those ten-dimensional solutions  will fall outside the classification of supersymmetric $AdS_3$ vacua of  type IIB supergravity given in \cite{Kim:2005ez,Gauntlett:2007ts}. The reason is that the $AdS_3$ solutions in \cite{Kim:2005ez,Gauntlett:2007ts} are  supported only by $F_{(5)}$ flux, while 
the Pilch-Warner solution of type IIB supergravity \cite{Pilch:2000ej}, which is the uplift of the KPW point, involves  non-trivial $C_{(3)}$, $H_{(3)}$ and $F_{(5)}$ fluxes. We expect that the same will hold  for the uplifts of the $AdS_3$ vacua in Section~\ref{Sec:AdS3sol}.

It should be possible to recast our $AdS_3 \times \Sigma_{\frak{g}}$ solutions in terms of three-dimensional $\mathcal{N}=2$ gauged supergravity along the lines of \cite{Karndumri:2013iqa,Karndumri:2013dca}.  It is  clear that the function $\mathcal{V}$ defined in Section~\ref{Sec:AdS3sol} should have some interesting interpretation. To this end one can reduce the action \eqref{CCaction} on $\Sigma_g$ and show that the three-dimensional effective gravitational action has the following potential
\begin{equation}
\mathcal{P}_{3D} = e^{-4h}\mathcal{P} + \frac{e^{-8h}}{2g^4} \left[\left(\frac{1}{4}+\mathfrak{a}\right)^2 e^{4\alpha-4\beta}+\left(\frac{1}{4}-\mathfrak{a}\right)^2 e^{4\alpha+4\beta}+\frac{1}{4}e^{-8\alpha}\right]+ \frac{e^{-6h}}{2g^2}\;.
\end{equation}
One then finds that
\begin{equation}
\mathcal{P}_{3D} =\frac{1}{4g^2}\left[\ds\sum_{i}(\partial_{\varphi_i}\mathcal{V})^2-2\mathcal{V}^2\right]\;,
\end{equation}
where the sum runs over the fields $i=\alpha,\beta,\chi,h$. Thus one can think of $\mathcal{V}$ as a superpotential for some three-dimensional gauged supergravity and this explains why the critical points of $\mathcal{V}$ are precisely the $AdS_3$ vacua we found in Section \ref{Sec:AdS3sol}. This structure is very similar to the one observed in \cite{Karndumri:2013iqa,Karndumri:2013dca}. 

In addition to the flows between fixed points discussed in Section~\ref{Sec:HoloRG}, there are also solutions in supergravity  that start at an $AdS_5$ or $AdS_3\times \Sigma_{\frak{g}}$ fixed point in the UV and diverge in the IR. Those are similar to the ``flows to Hades''  in conventional holographic RG flows \cite{Freedman:1999gp,Gubser:2000nd}. It would be interesting to identify
some criterion, along the lines of \cite{Gubser:2000nd,Maldacena:2000mw}, that would distinguish which of these flows are physical from the point of view of the dual field theory. 

Finally, using a similar approach as in this paper, one should be able to construct supersymmetric $AdS_2\times \Sigma_{\frak{g}}$ solutions in four-dimensional supergravity dual to  twisted compactifications on $\Sigma_g$ of the three-dimensional $\mathcal{N}=2$ SCFT \cite{Benna:2008zy}, which is a mass-deformed  ABJM theory. There should be also  holographic RG flows from the $AdS_4$ CPW solution \cite{Corrado:2001nv} to those new $AdS_2\times \Sigma_{\frak{g}}$  solutions.

\bigskip
\bigskip
\leftline{\bf Acknowledgements}
\smallskip
NB would like to thank Francesco Benini and Marcos Crichigno for many discussions and collaboration on \cite{BBC}. NB is also grateful to Mathew Bullimore for useful discussions. We would like to thank Anna Ceresole and Gianguido Dall'Agata for correspondence and Francesco Benini for comments on a draft of this paper. The work of NB is supported by Perimeter Institute for
Theoretical Physics. Research at Perimeter Institute is supported by the Government of Canada through Industry Canada and by the Province of Ontario through the Ministry of Research and Innovation. The work of KP and OV is supported in part by DOE grant DE-FG03-84ER-40168. OV would also like to thank the USC Dana and David Dornsife College of Letters, Arts and Sciences for support through the College Doctoral Fellowship and the CEA-Saclay for hospitality while this work was being completed.

\begin{appendices}

\appendix
\section{A truncation  of  $\cN=8$, $d=5$ supergravity}

\renewcommand{\theequation}{A.\arabic{equation}}
\renewcommand{\thetable}{A.\arabic{table}}
\setcounter{equation}{0}
\label{appendixA}

In this appendix we derive the  bosonic  action for the $U(1)_R$-invariant sector of  $\cN=8$, $d=5$ supergravity. Further truncation to the $U(1)_F$-invariant subsector and a trivial dilaton/axion  gives then the action  \eqref{CCaction}. 

First, we would like to clarify  a subtle point that otherwise might be confusing. It  follows from \eqref{branch8}  that by imposing the same symmetries on the fermionic fields one obtains a consistent truncation of $\cN=8$, $d=5$  supergravity  to some  $\cN=2$,~$d=5$   supergravity. In fact, we will find it convenient  to present  our results below in the language of    $\cN=2$, $d=5$ gauged  supergravity  \cite{Ceresole:2000jd}. However, since the Killing spinors for unbroken supersymmetries in   Section~\ref{Sec:SUGRAtrunc} are charged under $U(1)_R$,  the BPS equations that we obtain and solve are {\it not} for that $\cN=2$, $d=5$ supergravity. Instead, our analysis in Section~\ref{Sec:SUGRAtrunc} is carried out in the full $\cN=8$, $d=5$ supergravity. The truncation in the bosonic sector allows us to identify the fields that can be nontrivial and is crucial in making the entire analysis of the supersymmetry variations in Section~\ref{Sec:SUGRAtrunc} managable.

Recall that the  Lie algebra of $E_{6(6)}$ in the $SL(6,\RR)\times SL(2,\RR)$ basis (see, \cite{Gunaydin:1985cu}) consists of the $SL(6,\RR)$ generators, $\Lambda ^I{}_J$, $SL(2,\RR)$ generators, $\Lambda^\alpha{}_\beta$, and the generators $\Sigma_{IJK\alpha}$, where $I,J,\ldots=1,\ldots,6$ are the $SO(6)$ vector indices, while $\alpha,\beta=1,2$ are the vector indices of $SL(2,\RR)$. The 42 noncompact generators comprise of  the traceless $\Lambda^I{}_J\eql \Lambda ^J{}_I$, the self-dual tensors, $\Sigma_{IJK\alpha}$, and the traceless $\Lambda^\alpha{}_\beta=\Lambda^\beta{}_\alpha$ that transform in $\bf 20'$,  $\bf 10\oplus \overline{10}$ and $\bf 1\oplus 1$ of $SO(6)$, respectively.

\def\po{{\phantom{-}1}}
The $U(1)_R\subset SO(6)$ symmetry generator is 
\begin{equation}\label{Tigens}
T_R\eql {1\over 2} \left(\begin{matrix}
T_1 & &\\
& T_2 & \\
& & 2 \,T_3
\end{matrix}\right)\,,
\end{equation}
 where $T_i$, $i=1,2,3$, are the three $SO(2)$ generators,
\begin{equation}\label{so2s}
T_i\eql \left(\begin{matrix}
\phantom{-}0 & 1 \\ -1 &  0
\end{matrix}\right)\,,
\end{equation}
in the Cartan subalgebra of $SO(6)$.

The bosonic  fields  of $\cN=8$, $d=5$ supergravity that are invariant under $U(1)_R$ 
and comprise the bosonic sector of the corresponding  $\cN=2$, $d=5$ gauged supergravity are  the graviton, $g_{\mu\nu}$,    five vector fields for the commutant of $U(1)_R$ in $SO(6)$ and eight scalar fields  arising from the noncompact generators in $E_{6(6)}$ that commute with $U(1)_R$.\footnote{The two-form fields,   $B_{\mu\nu}^{I\alpha}$,  of the $\cN=8$  supergravity are all charged under $U(1)_R$ and hence they do not contribute to the truncation. }

The invariant vector fields are given explicitly by 
\begin{equation}\label{ungagef}
A^{(a)}\,T_{a}\eql 
\left(\begin{matrix}
  0 & A^{ {(1)}} & A^{ {(4)}} & A^{ {(5)}} & 0 & 0 \\
 -A^{ {(1)}} & 0 & -A^{ {(5)}} & A^{ {(4)}} & 0 & 0 \\
 -A^{ {(4)}} & A^{ {(5)}} & 0 & A^{ {(2)}} & 0 & 0 \\
 -A^{ {(5)}} & -A^{ {(4)}} & -A^{ {(2)}} & 0 & 0 & 0 \\
 0 & 0 & 0 & 0 & 0 & A^{ {(3)}} \\
 0 & 0 & 0 & 0 & -A^{ {(3)}} & 0 
\end{matrix}\right)\,.
\end{equation}
In particular, the fields $A^{(1)}$, $A^{(2)}$ and $A^{(3)}$ are the ones considered in Section \ref{Sec:SUGRAtrunc}. 
With all fields present, the unbroken gauge symmetry is  $SU(2)_\ell\times U(1)_R\times U(1)$, with the corresponding generators given by the following linear combinations of the generators in \eqref{ungagef}:
\begin{equation}\label{curlyTs}
\begin{gathered}
\cals T_1 \eql T_5\,,\qquad \cals T_2\eql - T_4\,,\qquad \cals T_3\eql T_1-T_2\,,\\[4 pt]
\cals T_4 \eql T_1+T_2+2\, T_3\,,\qquad \cals T_5\eql T_1+T_2-T_3\,,
\end{gathered}
\end{equation}
As one might have expected, the structure of the resulting truncation becomes more transparent when written in terms of gauge fields, $\cA^a$, $a=1,\ldots,5$,  with respect to this basis. Setting
\begin{equation}\label{}
 \cA^a\,\cT_a\eql A^{(a)}T_a \,,
\end{equation}
we find
\begin{equation}\label{curlytoflatA}
\begin{gathered}
\mathcal{A}^{1}  \eql A^{(5)}\;, \qquad \mathcal{A}^{2} \eql -A^{(4)}\;, \qquad \mathcal{A}^{3} = \frac{1}{2}(A^{(1)}-A^{(2)})\;,\\[4 pt]
\mathcal{A}^{4}  \eql \frac{1}{6}(A^{(1)}+A^{(2)}+2A^{(3)})\;, \qquad \mathcal{A}^{5}\eql \frac{1}{3}(A^{(1)}+A^{(2)}-A^{(3)})\;.
\end{gathered}
\end{equation}

The eight scalar fields, $\alpha$, $\beta^1,\,\beta^2,\,\beta^3$ and $w^1=x^1+i y^1$, $w^2=x^2+i y^2$,  parametrize a product of three noncompact coset spaces
\begin{equation}\label{slcoset}
\mathcal{M} = \mathcal{M}_{VS}\times \mathcal{M}_{QK} = \left[O(1,1)\times\dfrac{SO(3,1)}{SO(3)}  \right] \times \left[\dfrac{SU(2,1)}{SU(2)\times U(1)}\right]\;,
\end{equation}
where  $\mathcal{M}_{VS}$ is a very special manifold for the scalars in the vector multiplet and $\mathcal{M}_{QK}$ is a quaternionic K\"ahler manifold for the four real scalars in a full  five-dimensional $\cN=2$ hypermultiplet.  The scalars in $\mathcal{M}_{VS}$ arise from $\bf{20}'$ of $SO(6)$ in the $\cN=8$ theory. Two of the scalars in $\mathcal{M}_{SK}$ are the five-dimensional dilaton/axion, while the other two lie  in  $\bf{10}\oplus \overline{\bf{10}}$ of $SO(6)$. The four gauge fields in the vector multiplet gauge the $SO(3)$ and $U(1)$ isometries of $\mathcal{M}_{VS}$ and $\mathcal{M}_{QK}$, respectively, while  the  graviphoton of the five-dimensional theory is the gauge field for $U(1)_R$. This agrees nicely with the fact that the graviphoton should be dual to the superconformal $R$-symmetry of the LS fixed point.
The kinetic term  in the truncated supergravity is a sum of three terms, one for each factor in \eqref{slcoset}. Let us discuss them in turn introducing the scalar fields along the way.
\smallskip

  The noncompact generator for the first factor in \eqref{slcoset} is the diagonal element in $SL(6,\RR)$, 
\begin{equation}\label{}
X^{(\alpha)}\eql \mathop{\text{diag}}(-1,-1,-1,-1,2,2)\,,
\end{equation}
with the corresponding  $O(1,1)\subset E_{6(6)}$ group elements
\begin{equation}\label{}
V(\alpha)\eql \exp(\alpha \,X^{(\alpha)})\,.
\end{equation}
The kinetic term for  $\alpha$  is simply
\begin{equation}\label{alphalag}
e^{-1}\cL_{kin,\alpha}\eql 3\,g^{\mu\nu}\partial_\mu\alpha\,\partial_\nu\alpha\,.
\end{equation}

  The second coset in \eqref{slcoset} also arises from $U(1)_R$ invariant $SL(6,\RR)$ generators in $\bf 20'$ of $SO(6)$. The $SO(3,1)\subset E_{6(6)}$ group elements parametrizing this coset are
\begin{equation}\label{so31grp}
V(\beta^1,\beta^2,\beta^3)\eql \exp(\beta^1\,X^{(\beta)}_1+\beta^2\,X^{(\beta)}_2+\beta^3\,X^{(\beta)}_3)
\,,
\end{equation}
where
\begin{equation}\label{}
\beta^1\,X^{(\beta)}_1+\beta^2\,X^{(\beta)}_2+\beta^3\,X^{(\beta)}_3\eql 
\left(\begin{matrix}
\beta ^3 & 0 & \beta ^1 & \beta ^2 & 0 & 0 \\
 0 & \beta ^3 & -\beta ^2 & \beta ^1 & 0 & 0 \\
 \beta ^1 & -\beta ^2 & -\beta ^3 & 0 & 0 & 0 \\
 \beta ^2 & \beta ^1 & 0 & -\beta ^3 & 0 & 0 \\
 0 & 0 & 0 & 0 & 0 & 0 \\
 0 & 0 & 0 & 0 & 0 & 0 \\
\end{matrix}\right)\,.
\end{equation}
The normalization of the fields and generators here has been chosen to agree with the truncation in Section \ref{Sec:SUGRAtrunc}. The group elements \eqref{so31grp} are then isomorphic with 
\begin{equation}\label{}
g(\beta^1,\beta^2,\beta^3)\eql \exp\,\left(\begin{matrix}
0 & 0 & 0 & 2\beta^1 \\
0 & 0 & 0 & 2\beta^2 \\
0 & 0 & 0 & 2\beta^3 \\
2\beta^1 & 2\beta^2 & 2\beta^3 & 0 \\
\end{matrix}\right)\,,
\end{equation}
in the fundamental representation of $SO(3,1)$. Thus the standard projective coordinates on this  coset   are given by 
\begin{equation}\label{}
 \xxi^i\eql {\beta^i\over \bfs\beta}\tanh 2\bfs \beta \,,\qquad \bfs \beta\eql \sqrt{(\beta^1)^2+(\beta^2)^2+(\beta^3)^2}\,.
\end{equation}
In terms of these coordinates,  the kinetic action in this sector is
\begin{equation}\label{zlag}
e^{-1}\cals L_ {kin,\xxi}\eql {1\over 4}\,g^{\mu\nu} \,\Big[\, {\nabla_\mu \xxi^i\nabla_\nu \xxi^i\over 1- \bfs\xxi{}^2}+ {( \xxi^i\nabla_\mu \xxi^i)( \xxi^j\nabla_\nu \xxi^j)\over ( 1- \bfs\xxi{}^2)^2}\,
\Big]\,,
\end{equation}
where $\bfs\xxi{}^2=( \xxi^1)^2+( \xxi^2)^2+( \xxi^3)^2$. The gauge covariant derivative is
\begin{equation}\label{}
\nabla \xxi^i\eql d \xxi^i+g\,\mathcal{A}^{a}\mathcal{K}_{(a)}^z{}^{i} \,,
\end{equation}
with the Killing vector fields
\begin{equation}\label{killz}
\begin{gathered}
\mathcal{K}_{(1)}^z   \eql (0, 2\, \xxi^3,-2\, \xxi^2)\,, \qquad \mathcal{K}_{(2)}^z  \eql (-2\, \xxi^3,0,2\, \xxi^1)\,, \qquad \mathcal{K}_{(3)}^z   \eql (2\, \xxi^2,-2\, \xxi^1,0)\,,\\[6 pt]
\mathcal{K}_{(4)}^z    \eql \mathcal{K}_{(5)}^z  \eql (0,0,0) \,.
\end{gathered}
\end{equation}
Note that only the $SU(2)_\ell$ symmetry is gauged by this sector.  

The last coset in \eqref{slcoset} is the most intricate and can be parametrized in a variety of ways.\footnote{For an extensive discussion and earlier references, see \cite{Ceresole:2001wi,Suh:2011xc}.} Here we will use two parametrizations, one with a familiar form of the kinetic term and the other one with a manifest independence of the scalar potential on the dilaton/axion field.

There are  four noncompact generators in this sector. The first two are the tensor generators $\Sigma_{IJK\alpha}=\Sigma_{[IJK]\alpha}$  in $\bf 10\oplus\overline{10}$ of $SO(6)$ and  the nonvanishing components:
\begin{equation}\label{}
\begin{split}
X^{(x)}_1~:\qquad \Sigma_{1361}& \eql -\Sigma_{1451}=-\Sigma_{2351}\eql-\Sigma_{2461}\\
 & \eql -\Sigma_{1352}\eql -\Sigma_{1462}\eql-\Sigma_{2362}\eql\Sigma_{2452}\eql {1\over 2\sqrt 2}\,,\\
X^{(y)}_1~:\qquad \Sigma_{1351} & \eql \Sigma_{1461} \eql \Sigma_{2361} \eql -\Sigma_{2451}\\ &
\eql \Sigma_{1362} \eql -\Sigma_{1452} \eql -\Sigma_{2352} \eql -\Sigma_{2462} \eql {1\over 2\sqrt 2}\,,
\end{split}
\end{equation}
while the other two are the dilaton/axion singlets in  $SL(2,\RR)$,
\begin{equation}\label{}
X^{(x)}_2\eql \left(\begin{matrix}
1 & \phantom{-}0\\ 0 & -1
\end{matrix}\right)\,,\qquad X^{(y)}_2\eql \left(\begin{matrix}
0 & 1\\ 1 & 0
\end{matrix}\right)\,.
\end{equation}
With those choices,   the group elements 
\begin{equation}\label{}
V(x^1,y^1,x^2,y^2)\eql\exp(x^1 X^{(x)}_1+y^1  X^{(y)}_1 + x^2 X^{(x)}_2 + y^2  X^{(y)}_2)\,,
\end{equation}
in $SO(2,1)\subset E_{6(6)}$ yield the same parametrization of the coset as the more familiar $SU(2,1)$ matrices,
\begin{equation}\label{}
g(w^1,w^2)\eql \exp\left(\begin{matrix}
0 & 0 & w^1\\
0 & 0 & w^2\\
\overline w{}^1 & \overline w{}^2 & 0
\end{matrix}\right)\,,\qquad w^j\eql x^j+i\,y^j\,.
\end{equation}
In terms of the projective coordinates,
\begin{equation}\label{}
\zeta^i\eql {w^i\over w}\tanh w\,,\qquad w^2\eql |w^1|^2+|w^2|^2\,,
\end{equation}
the kinetic part of the action in this sector has the familiar noncompact Fubini-Study  form
\begin{equation}\label{zetalag}
e^{-1}\cals L_{kin,\zeta} \eql {1\over 2}\,g^{\mu\nu}\left[{\nabla_\mu \zeta^i\nabla_\nu\bar\zeta{}^i\over 1-| \bfs\zeta|^2}
+{(\zeta^i\nabla_\mu\bar\zeta{}^i)(\bar\zeta{}^j\nabla_\nu\zeta{}^j)\over ( 1-| \bfs\zeta|^2){}^2}
\right]\,,
\end{equation}
where $|\bfs\zeta|^2=|\zeta^1|^2+|\zeta^2|^2$. The covariant derivatives   are
\begin{equation}\label{covdevzeta}
\nabla\zeta^i\eql d\zeta^i+g \,\cA^a\,\cK_{(a)}^\zeta{}^{i}\,,
\end{equation}
where
\begin{equation}\label{kilzeta}
\mathcal{K}_{(1)}^\zeta \eql \mathcal{K}_{(2)}^\zeta \eql  \mathcal{K}_{(3)}^\zeta  \eql \mathcal{K}_{(4)}^\zeta  \eql (0,0)\,,\qquad  \mathcal{K}_{(5)}^\zeta  \eql (3i\zeta^1,0)\,.
\end{equation}
This shows that only a single $U(1)$ is gauged in this sector.

The price that one pays for the simplicity of the kinetic action in this parametrization is that the scalar potential, when restricted to the fields in this sector, reads
\begin{equation}\label{potzeta}
\cals P_\zeta\eql -{3\over 8}{( 2-3|\zeta^1|^2-2|\zeta^2|^2)(1-|\zeta^2|^2)\over (1-|\zeta^1|^2-|\zeta^2|^2)^2}\,.
\end{equation}
and thus depends  on both $\zeta^1$ and $\zeta^2$. This makes the identification of the dilaton/axion fields somewhat tricky \cite{Clark:2005te}. 

Instead, one can use  another parametrization in which the coset is decomposed locally  as the product
\begin{equation}\label{splitpar}
{ SU(2,1)\over SU(2)\times U(1)}\simeq {SU(1,1)\over U(1)}\cdot {SU(1,1)\over U(1)}\,,
\end{equation}
with the complex fields $\xi^1$ and $\xi^2$, respectively, where $\xi^2$ is the dilaton/axion field. This new parametrization amounts to the field redefinition
\begin{equation}\label{chvar}
\zeta^1\eql \xi^1\,\sqrt{1-|\xi^2|^2}\,,\qquad \zeta^2\eql\xi^2\,,
\end{equation}
which can be applied to any (composite) gauge invariant expressions.\footnote{In general, there will be a compensating composite $SU(2)\times U(1)\subset USp(8)$ gauge transformation that must be performed on top of the field redefinition.} In particular, the kinetic action in terms of those fields is obtained from \eqref{zetalag} and involves covariant derivatives 
\begin{equation}\label{covdexi}
\nabla\xi^i\eql d\xi^i+g \,\cA^{(a)}\cK_{(a)}^\xi{}^{i}\,,
\end{equation}
with the Killing vectors
\begin{equation}\label{kilxi}
\mathcal{K}_{(1)}^\xi \eql \mathcal{K}_{(2)}^\xi \eql  \mathcal{K}_{(3)}^\xi  \eql \mathcal{K}_{(4)}^\xi  \eql (0,0)\,,\qquad\qquad \mathcal{K}_{(5)}^\xi  \eql (3i\xi^1,0)\,.
\end{equation}
The potential \eqref{potzeta}  now becomes
\begin{equation}\label{}
\cals P_\xi \eql -{3\over 8}{( 2-3|\xi^1|^2)\over (1-|\xi^1|^2}\,.
\end{equation}
and is  manifestly  independent of the dilaton/axion field, $\xi^2$.

Using this parametrization, it is now straightforward to calculate the full scalar potential with all eight scalar fields.  It reads
\begin{equation}\label{finpot}
\begin{split}
\cals P & \eql {1\over 8}\,e^{8\alpha}\,{|\xi^1|^2\over (1-|\xi^1|^2)^2}-{1\over 2}\,e^{2\alpha}\,{1\over \sqrt{1-\bfs z^2}\, (1-|\xi^1|^2) } - {1\over 4}\,e^{-4\alpha}\,{1-\bfs z^2-2\,|\xi^1|^2\over (1-\bfs z^2)\, (1-|\xi^1|^2)^2 }\,.
\end{split}
\end{equation}
It does not depend on the dilaton/axion and  is invariant under the gauge transformations generated by  \eqref{killz} and \eqref{kilxi}.

Let us now turn to the vector fields.  We find that the Maxwell action reduces to 
\begin{equation}\label{maxact}
e^{-1}\cals L_{Max}\eql -{1\over 4}\,a_{ab}\,\mathcal{F}^{(a)}_{\mu\nu}\mathcal{F}^{(b)}{}^{\mu\nu}\,,
\end{equation}
with the field strengths
\begin{equation}
\cals F^a\eql d\,\cals A^a-{g\over 2}\,f_{bc}{}^a\,\cals A^b\wedge \cals A^c\,,
\end{equation}
where $f_{ab}{}^c$ are the structure constants,   $[\cals T_a,\cals T_b]=f_{ab}{}^c \cals T_c$.
The matrix, $(a_{ab})$,  $a,b=1,\ldots,5$, of Yang-Mills couplings  is given explicitly by 
\begin{equation}\label{}
\begin{split}
a_{ij} & \eql 2\,e^{4\alpha}\, \Big(\delta^{ij}+2 \,{z^iz^j\over 1-\bfs z^2}\Big)\,,\qquad\qquad a_{i4}  \eql  -4\,e^{4\alpha}\,{z^i\over 1-\bfs z^2}\,,  \qquad\qquad i,j=1,2,3\,,
\\[5pt]
a_{44}&\eql 4\,e^{-8\alpha}+2\,e^{4\alpha}\,{1+\bfs z^2\over 1-\bfs z^2}\,,~~ 
a_{45}  \eql -2\,e^{-8\alpha}+2\,e^{4\alpha}\,{1+\bfs z^2\over 1-\bfs z^2}\,,~~
a_{55}  \eql e^{-8\alpha}+2\,e^{4\alpha}\,{1+\bfs z^2\over 1-\bfs z^2}\,.
\end{split}
\end{equation}
It depends only on the scalar fields $\alpha$ and $z^1,\ldots,z^3$.

The full action of the truncated theory is thus 
\begin{equation}\label{fulllag}
\cals L\eql -{1\over 4}\,e\,R+\cals L_{Max}+\cals L_{kin,\alpha}+\cals L_{kin,z}+\cals L_{kin,\zeta}-e\,g^2\,\cals P+\cals L_{CS}\,,
\end{equation}
with the individual terms given in \eqref{maxact}, \eqref{alphalag}, \eqref{zlag}, \eqref{zetalag} and \eqref{finpot}, respectively. The last term in \eqref{fulllag} is the Chern-Simons term which we discuss in more detail below.

The $U(1)_F$-invariant subtruncation is now simply obtained by setting
\begin{equation}\label{}
\beta^1\eql \beta^2\eql 0\,,\qquad \beta^3\eql \beta\qquad \text{or}\qquad z^1\eql z^2\eql 0\,,\qquad z^3\eql \tanh 2\beta\,,
\end{equation}
which reduces the coset \eqref{slcoset} to \eqref{slcosetM}, 
and restricting the vector fields to the Abelian subalgebra, $\cals A^1=\cals A^2=0$.
Finally, the truncation in Section~\ref{Sec:SUGRAtrunc} is obtained by turning off the dilaton and axion field, 
\begin{equation}\label{}
w^1\eql i\,\chi\,e^{i\theta}\,,\qquad w^2\eql 0\qquad \text{or}\qquad  \zeta^1\eql \xi^1\eql i\,\tanh\chi\,e^{i\theta}\,,\qquad \zeta^2\eql\xi^2\eql 0\,,
\end{equation}
One can verify that at the level of equations of motion the latter is a consistent truncation.

Let us conclude with some comments. First, one may wish to check that the   $U(1)_R$-invariant truncation of the fermionic sector indeed yields the correct fields to complement the bosonic sector of the putatitive $\cN=2$, $d=5$ gauged supergravity. We have already seen that out of the eight gravitini of the maximal theory, two are invariant under $U(1)_R$, see \eqref{branch8}. They correspond to the two gravitini in the five-dimensional $\mathcal{N}=2$ gravity multiplet. Similarly, out of the forty eight spin-1/2 fields in the $\mathcal{N}=8$ theory, there are ten singlets under $U(1)_R$, out of which eight belong to the four vector multiplets, and two are in the hypermultiplet. Given the bosonic action in \eqref{fulllag}, we know both the precise scalar coset \eqref{slcoset} and the Killing vectors \eqref{killz}, \eqref{kilxi} of the symmetries that are gauged.  In Appendix~\ref{appendixB}, we will also obtain  the Chern-Simons couplings. With this information at hand it should be possible to recover the full  $\mathcal{N}=2$, $d=5$  gauged supergravity \cite{Ceresole:2000jd} corresponding to this truncation.

Secondly, one can calculate  the linearized mass spectrum of the eight scalars in this truncation around the $AdS_5$ critical points \eqref{SO6fp}-\eqref{SU3fp} and determine the dimensions of the dual operators using the standard relation $m^2L^2=\Delta(\Delta-4)$. This provides a useful comparison with the known results for the full $\cN=8$ theory \cite{Kim:1985ez,Distler:1998gb,Freedman:1999gp}. Around the $SO(6)$ critical point \eqref{SO6fp}, one finds that the axion-dilaton has $m^2L^2=0$, which is appropriate for the marginal complexified YM coupling of $\mathcal{N}=4$ SYM. The scalars $\alpha$ and $\beta_i$ have $m^2L^2=-4$ and are dual to bosonic bilinear operators with $\Delta=2$. The remaining two scalars in the hypemultiplet have $m^2L^2=-3$ and are dual to fermionic bilinear operators of dimension $\Delta=3$. Around the KPW critical point \eqref{PWfp}, the axion-dilaton together with the field $\theta$ have $m^2L^2=0$ and are thus dual to marginal operators in the LS SCFT. The triplet of scalars $\beta_i$ all have $m^2L^2=-4$ and are dual to operators with $\Delta=2$. These operators belong to the short multiplet containing the conserved $SU(2)_F$ current (see, Table~6.1 in \cite{Freedman:1999gp}). The scalars $\alpha$ and $\chi$ mix at the linearized level and the eigenvalues of the mass matrix are $m^2L^2=2(2\pm \sqrt{7})$, which corresponds to a relevant operator of dimension $\Delta=1+\sqrt{7}$ and an irrelevant one of dimension $\Delta=3+\sqrt{7}$. These two operators belong to the unprotected multiplet in the first entry of Table~6.2 in \cite{Freedman:1999gp}. Finally, at the   $SU(3)$ invariant critical point \eqref{SU3fp}, the scalar $\chi$ has $m^2L^2=8$ and all other seven scalars have $m^2L^2=0$. Since this point is perturbatively unstable in the full $\mathcal{N}=8$ theory, it is unclear whether one can interpret it holographically.

\section{Chern-Simons levels and anomalies}

\renewcommand{\theequation}{B.\arabic{equation}}
\renewcommand{\thetable}{B.\arabic{table}}
\setcounter{equation}{0}
\label{appendixB}

The Chern-Simons term of the $U(1)_R$-invariant truncation discussed in  Appendix~\ref{appendixA} can be read off from \cite{Gunaydin:1985cu} 
\begin{equation}
\mathcal{L}_{CS} = C_{abc} \,\Big[\mathcal{F}^{a}\wedge \mathcal{F}^{b} \wedge \mathcal{A}^{c} - \frac{3g}{4}f_{de}{}^{c} \mathcal{F}^{a} \wedge \mathcal{A}^{b}\wedge \mathcal{A}^{d} \wedge \mathcal{A}^{e}  \Big]\;,
\end{equation}
where $f_{ab}{}^c$ are the structure constants of the gauge group.\footnote{The $\mathcal{A}\wedge \mathcal{A}\wedge \mathcal{A}\wedge \mathcal{A}\wedge \mathcal{A}$ CS term present in the $\mathcal{N}=8$ theory is identically zero in this truncation.} 
The symmetric tensor of CS couplings, $C_{abc}$, has only nine non-zero components given by
\begin{equation}
\begin{gathered}
C_{114} = C_{224}  = C_{334} = \frac{4}{3} \;,  \qquad C_{115} = C_{225}  = C_{335} = -\frac{2}{3} \;, \\
C_{444} = -4\;,  \qquad C_{445} = -2\;,  \qquad C_{555} = 2\;.
\end{gathered}
\end{equation}

In $\mathcal{N}=4$ SYM the supergravity gauge fields $\mathcal{A}^a$ correspond to conserved currents $\mathcal{J}^a$. One can then define the matrix of 't Hooft anomalies for the global currents $\mathcal{J}^a$ as
\begin{equation}
k^{abc} = \text{Tr}(\mathcal{J}^a\mathcal{J}^b\mathcal{J}^c)\,,
\end{equation}
where the trace is to be taken over all fermions in $\mathcal{N}=4$ SYM. Using the charges of the fermions in Table~\ref{table:one}, one finds that the non-vanishing components of $k^{abc}$ are
\begin{equation}
\begin{gathered}
k^{114}  = k^{224}  = k^{334} = -\frac{1}{4}d_{G} \;,  \qquad k^{115} = k^{225}  = k^{335} = \frac{1}{8}d_{G} \;, \\
k^{444}  = \frac{3}{4}d_{G} \;,  \qquad k^{445} = \frac{3}{8}d_{G}\;, \qquad k^{555} = -\frac{3}{8}d_{G}\;.
\end{gathered}
\end{equation}
It is a consequence of holography that the matrix of Chern-Simons couplings for gauge fields in five-dimensional supergravity is proportional to the matrix of 't Hooft anomalies for global currents in the dual field theory. Indeed we find
\begin{equation}
k^{abc} = -\frac{3 d_G}{16}\,C_{abc}\;.
\end{equation}
It is worth noting that at the LS fixed point the conserved current $\mathcal{J}^5$ is not present. This is manifested in supergravity by the gauge field $\mathcal{A}^5$ becoming massive at the KPW point due to the non-zero value of the scalar field $\chi$.

At the $AdS_3$ vacua of interest, the scalar $\beta$ is generically nonzero which breaks the $SU(2)$ gauge symmetry by giving mass to the $\mathcal{A}^{1}$  and $\mathcal{A}^{2}$ gauge fields. Thus we are left with two massless gauge fields $\mathcal{A}^{3}$  and $\mathcal{A}^{4}$ corresponding to the two Abelian global symmetries of the $(0,2)$ SCFTs in the IR. The value $\mathfrak{a}=0$ is special because   $\beta=0$ and we preserve the full $SU(2)$ gauge symmetry in the gravity theory. Then the dual $(0,2)$ SCFT  has $SU(2)$ flavor symmetry in addition to the omnipresent $U(1)_R$ symmetry.

The two supergravity gauge fields that are massless at the $AdS_3\times \Sigma_{\frak g}$ solutions are
\begin{equation}
\mathcal{F}^3 = \dfrac{(a_1-a_2)}{2} \frac{dx\wedge dy}{y^2} = \frac{\mathfrak{a}}{g} \frac{dx\wedge dy}{y^2}\;, \qquad \mathcal{F}^4 = \dfrac{(a_1+a_2+2a_3)}{6} \frac{dx\wedge dy}{y^2}= \frac{1}{4g} \frac{dx\wedge dy}{y^2}\;,
\end{equation}
where   we have used the explicit change of basis  \eqref{curlytoflatA}.
Upon dimensional reduction to three dimensions, one finds the following Chern-Simons term
\begin{equation}
\mathcal{L}^{(3)}_{CS} =C_{ab} \,\mathcal{F}^{a} \wedge \mathcal{A}^{b} \;,
\end{equation}
where the symmetric matrix of Chern-Simons couplings is
\begin{equation}
C_{33} = \frac{1}{g} 2\pi\eta_{\Sigma}\;, \qquad C_{34} =C_{43}= \frac{4\mathfrak{a}}{g} 2\pi\eta_{\Sigma}\;, \qquad  C_{44}= - \frac{3}{g} 2\pi\eta_{\Sigma}\;,
\end{equation}
and $\eta_{\Sigma}$ is defined below \eqref{nrnldef}.

The matrix of two-dimensional current anomalies is 
\begin{equation}\label{kIJ}
k^{ab} = -\eta_{\Sigma} d_{G} \left[ t_{\frak{b}}^{(\lambda)}q_{a}^{(\lambda)}q_{b}^{(\lambda)} +t_{\frak{b}}^{(\chi_2)} q_{a}^{(\chi_2)}q_{b}^{(\chi_2)} + t_{\frak{b}}^{(\chi_3)}q_{a}^{(\chi_3)}q_{b}^{(\chi_3)} \right]\;,
\end{equation}
where the charges of the fermion fields can be read off from Table \ref{table:two} with the index $a,b=3$ corresponding to $U(1)_F$ and $a,b=4$ to $U(1)_R$, respectively.
A short calculation yields
\begin{equation}
k^{33} = - \frac{1}{8}d_G\eta_{\Sigma}\;, \qquad k^{34} = k^{43} = \frac{\mathfrak{b}}{4}d_G\eta_{\Sigma}\;, \qquad k^{44} =  \frac{3}{8}d_G\eta_{\Sigma}\;.
\end{equation}
Thus we arrive at the expected result that the matrix of Chern-Simons level in the three-dimensional gravitational theory is proportional to the matrix of current anomalies in the dual CFT
\begin{equation}
k^{ab} =  -\frac{g d_G}{16\pi}\,C_{ab}\,.
\end{equation}
This holds provided we set $\mathfrak{b}=-2\mathfrak{a}$, which agrees with the discussion below \eqref{ccsugra}.  

\end{appendices}


\end{document}